\documentclass[]{aa}

\usepackage{graphicx}
\usepackage{adjustbox}
\usepackage{txfonts}
\usepackage{amsmath}
\usepackage[nopatch]{microtype}
\usepackage{booktabs}
\usepackage{mathtools}

\usepackage{hyperref}
\hypersetup{
    colorlinks=true,
    linkcolor=blue,
    filecolor=magenta,      
    urlcolor=blue,
    citecolor=blue,
    }
\begin{document} 
\title{A $JWST$ project on 47~Tucanae. Binaries among multiple populations.}
\subtitle{}

\author{A.\,P.\,Milone\inst{1,2},
A.\,F.\,Marino\inst{2},
M.\,Bernizzoni\inst{1},
F.\,Muratore\inst{1},
M.\,V.\,Legnardi\inst{1},
M.\,Barbieri\inst{1},
E.\,Bortolan\inst{1},
A.\,Bouras\inst{1},
J.\,Bruce\inst{3},
G.\,Cordoni\inst{4},
F.\,D'Antona\inst{5},
F.\,Dell'Agli\inst{5},
E.\,Dondoglio\inst{2},
I.\,M.\,Grimaldi\inst{1},
S.\,Jang\inst{6},
E.\,P.\,Lagioia\inst{7},
J.\,-W.\,Lee\inst{8},
S.\,Lionetto\inst{1},
A.\,Mohandasan\inst{1},
X.\,Pang\inst{9,10},
C.\,Pianta\inst{1},
M.\,Posenato\inst{1},
A.\,Renzini\inst{2},
M.\,Tailo\inst{2},
C.\,Ventura\inst{5},
P.\,Ventura\inst{5},
E.\,Vesperini\inst{3},
T.\,Ziliotto\inst{1},
}


\institute{Dipartimento di Fisica e Astronomia ``Galileo Galilei'', Univ. di Padova, Vicolo dell'Osservatorio 3, Padova, IT-35122 \\ \email{antonino.milone@unipd.it}
\and
Istituto Nazionale di Astrofisica - Osservatorio Astronomico di Padova, Vicolo dell’Osservatorio 5, Padova, IT-35122
\and
Department of Astronomy, Indiana University, Bloomington, Swain West, 727 E. 3rd Street, IN 47405, USA
\and 
Research School of Astronomy and Astrophysics, Australian National University, Canberra, ACT 2611, Australia
     \and 
     Istituto Nazionale di Astrofisica, Osservatorio Astronomico di Roma, Via Frascati 33, 00077 Monte Porzio Catone, Italy
\and
Center for Galaxy Evolution Research and Department of Astronomy, Yonsei University, Seoul 03722, Korea
\and
South-Western Institute for Astronomy Research, Yunnan University, Kunming, 650500 P. R. China
\and
Department of Physics and Astronomy, Sejong University, 209 Neungdong-ro, Gwangjin-Gu, Seoul 05006, Republic of Korea 
\and
Department of Physics, Xi'an Jiaotong-Liverpool University, 111 Ren'ai Road, Dushu Lake Science and Education Innovation District, Suzhou 215123, Jiangsu Province, People's Republic of China
\and
Shanghai Key Laboratory for Astrophysics, Shanghai Normal University, 100 Guilin Road, Shanghai 200234, People's Republic of China
}

\titlerunning{Binaries among the multiple populations of 47\,Tucanae} 
\authorrunning{A.\,P.\,Milone et al.}

\date{Received XXXX; accepted XXXX}

\abstract{Almost all globular clusters (GCs) contain multiple stellar populations consisting of stars with varying helium and light-element abundances. These populations include first-population stars, which exhibit similar chemical compositions to halo-field stars with comparable [Fe/H], and second-population stars, characterized by enhanced helium and nitrogen abundances along with reduced levels of oxygen and carbon.
 Nowadays, one of the most intriguing open questions about GCs pertains to the formation and evolution of their multiple populations. 

Recent works based on N-body simulations of GCs show that the fractions and characteristics of binary stars can serve as dynamic indicators of the formation period of multiple-population in GCs and their subsequent dynamical evolution. Nevertheless, the incidence of binaries among multiple populations is still poorly studied. Moreover, the few available observational studies are focused only on the bright stars of a few GCs.

In this work, we use deep images of the GC 47\,Tucanae collected with the James Webb and the Hubble space telescopes to investigate the incidence of binaries among multiple populations of M-dwarfs and bright main-sequence stars.
To reach this objective, we use UV, optical, and near infrared filters to construct photometric diagrams that allow us to disentangle binary systems and multiple populations. Moreover, we compared these observations with a large sample of simulated binaries.

In the cluster central regions, the incidence of binaries among first-population stars is only slightly higher than that of second-population stars.
 In contrast, in the external regions, the majority ($\gtrsim 85 \%$) of the studied binaries are composed of first population stars. Results are consistent with the GC formation scenarios where the second-population stars  originate in the cluster's central region, forming a compact and dense stellar group within a more extended system of first-population stars.}

\maketitle
%
\section{Introduction}
\label{sec:intro}
Galactic globular clusters (GCs) are among the most-ancient stellar systems in the Universe and serve as laboratories for testing theories of stellar structure
 and evolution. Their investigation is complementary to studies based on high redshift galaxies to shed light on the assembly of large scale structures 
 such as the Galactic halo \citep[e.g.\,][]{renzini2017a, dantona2023a}.
One of the most fascinating properties of GCs is that a large fraction of their stars exhibits distinctive chemical compositions that are rarely observed among field stars. These 'second-population' (2P) stars are enhanced in some light elements, (including helium, nitrogen, aluminum, and sodium) and depleted 
 in others (e.g.\,carbon and oxygen) with respect to the remaining GC stars, which are indicated as 'first population' (1P) and share the same 
content of light elements as $\alpha$-enhanced field stars with similar metallicity \citep[see][for reviews]{kraft1994a, gratton2004a, gratton2012a, gratton2019a, bastian2018a, milone2022a}.

 Although the origin of multiple populations in GCs is a widely debated topic \citep[][]{cottrell1981a, dantona1983a, dantona2016a, ventura2001a, decressin2007a, demink2009a, 
 bastian2013a, denissenkov2014a, renzini2015a, renzini2022a, gieles2018a, calura2019a, wang2020a, lacchin2022a, lacchin2024a}, most formation scenarios suggest that 2P stars
 formed in a high-density environment, in the innermost GC regions.
 The binary stars among the distinct stellar populations in GCs are efficient tools to test this hypothesis and constrain the cluster formation and dynamical evolution. 
 Indeed, as a consequence of the denser formation environment, the 2P binaries should be more strongly affected by stellar interactions and disrupted at larger rates than 1P binaries 
    \citep{vesperini2011a, hong2015a, hong2016a, hypki2022a}.  N-body simulations by Vesperini, Hong, and collaborators show that in multiple-population clusters, a denser 2P system embedded in a less concentrated 1P cluster can lead to similar binary fractions for 1P and 2P stars in the central regions, with 1P binaries making up the majority of the total binaries \citep[][]{hong2015a, hong2016a, milone2020a, bortolan2025a}. 

Early observational constraints on the incidence of binaries  among 1P and 2P stars of GCs rely on spectroscopy. 
\cite{dorazi2010a} detected five Barium-rich stars in five GCs and noticed that four of them have content of sodium and oxygen consistent with 1P stars, whereas one star alone belongs to the 2P. 
 The large barium abundance observed in these Ba-stars is linked to unseen evolved companions that transferred processed material to the visible star during their asymptotic giant branch phase \citep[e.g.,][]{mcclure1989a, luck1991a}. Consequently, the findings by D'Orazi and collaborators suggest a prevalence of binaries among 1P stars.

Similar conclusions come from works based on spectroscopy of binary stars identified by means of radial-velocity variations.
Based on a study involving 21 radial-velocity  binaries across ten GCs, \citet{lucatello2015a} concluded that the fraction of binaries among 1P stars is 4.1$\pm$1.7 times higher than the fraction observed among 2P stars. This conclusion aligns with the research by \citet{dalessandro2018a}, who determined that out of 12  radial-velocity binaries in NGC\,6362, only one belongs to the 2P. 

Most of the stars studied spectroscopically are red giant branch (RGB) stars located in the external cluster regions. The first photometric investigation of binaries in the cluster centers relies on multi-band images of bright main-sequence (MS) stars obtained with the {\it Hubble} Space Telescope ({\it HST}) and yields puzzling conclusions. In the core of M\,4, the 1P binaries incidence is about three times higher than the 2P incidence \citep{milone2020a} and a predominace of 1P binaries is also detected in the central region of NGC\,3201 \citep{marino2019a, kamann2020a}. In contrast, NGC\,288, NGC\,6352, NGC\,6362, NGC\,6397 and NGC\,6838 exhibit similar binary incidence rates in their central $\sim 2.7 \times 2.7$ square-arcmin regions \citep{milone2020a, milone2025a}. 

Recently, we initiated a project to explore faint stars in the GC 47\,Tucanae, by using spectra and images obtained through the {\it James Webb} Space Telescope ({\it JWST}) as part of the GO-2560 program (PI: Anna F.\,Marino) together with archive HST and JWST images. The initial findings from this project include the pioneering detection of the brown-dwarf cooling sequence within a GC \citep{marino2024a}, the identification of multiple stellar populations among M-dwarfs, and the first determination of their oxygen abundances by means of both photometry and spectroscopy \citep{milone2023a, marino2024a, marino2024b}.

 In this work, we study the incidence of binaries among 1P and 2P stars of 47\,Tucanae, which is a GC that hosts a small fraction of binaries that correspond to $\sim$1\% in the central regions \citep{milone2012b}. The paper is organized as follows. 
 Section\,\ref{sec:data} presents the dataset and summarizes the data reduction techniques. Section\,\ref{sec:binaries} illustrates the methods to study the binaries among the multiple populations of 47\,Tucanae. 
 The distribution of binaries in the pseudo two-color diagrams dubbed 'chromosome maps' (ChMs) is discussed in the Section\,\ref{sec:chm}, whereas  the Section\,\ref{sec:summary} is dedicated to the summary of the results and the conclusions.
 
\section{Observations and data reduction}
\label{sec:data}
To investigate the binaries among the multiple populations of 47 Tucanae, we analyzed stars in two different fields of view: a central field (RA$\sim 00^{\rm h}24^{\rm m}06^{\rm s}$, DEC$\sim -72^{\rm d}04^{\rm m}53^{\rm s}$) and an external field (RA$\sim 00^{\rm h}24^{\rm m}37^{\rm s}$, DEC$\sim -72^{\rm d}04^{\rm m}06^{\rm s}$) located at $\sim$7 arcmin west of the cluster center (about 2.2 $R_{\rm hl}$ where $R_{\rm hl}=3.17$ arcmin is the cluster half-light radius from the 2010 version of the \cite{harris1996a} catalogue).

 We used images of the central field collected through the F435W, F606W and F814W filters  of the Wide Field Channel of the Advanced Camera for Surveys (WFC/ACS) and images in the F275W, F343N, F336W, and F438W bands of the Ultraviolet and Visual Channel of the Wide Field Camera 3 (UVIS/WFC3) on board the {\it HST}. 

  For the external field, we investigated the M-dwarfs of 47\,Tucanae by using the photometric catalogs by \citet{marino2024a}, which are obtained from ACS/WFC data in the F606W and F814W bands and images collected through the F322W2 filter of the near-infrared camera (NIRCam) onboard the {\it JWST}. In addition, we used images collected with the F336W of UVIS/WFC3 and the F435W of WFC/ACS. These data, together with the photometry in the F606W and F814W bands from \cite{marino2024a} allowed us to explore the upper MS of 47\,Tucanae. The main properties of the images used to derive the photometry used in this paper are summarized in Table\,\ref{tab:data}.

Stellar photometry and astrometry are obtained using the computer program KS2, developed by Jay Anderson as an evolution of the  
 program initially developed to reduce ACS/WFC images from GO-10775 in the F606W and F814W bands \citep{anderson2008a}.
In summary, KS2 employs three methods to measure stars, optimizing astrometry and photometry for stars in different luminosity intervals. 
As our focus lies on relatively bright stars, we relied on the results obtained from method I. This method measures all stellar sources that produce a distinct peak within a 5$\times$5 pixel region after subtracting neighboring stars.

 To achieve this, KS2 calculates the flux and position of each star in each image separately by using the point spread function (PSF) model associated with its position. It also subtracts the sky level, which is computed from the annulus between 4 and 8 pixels from the stellar center.
 Finally, the results from all images are averaged to obtain the most accurate determinations of stellar magnitudes and positions. 
 The KS2 computer program offers multiple diagnostics for assessing photometric quality. To ensure high-precision photometry, we included only isolated stars that are well-fitted by the PSF model in our analysis \citep[see section 2.4 of][for details]{milone2023b}. Moreover, we excluded the region with a radial distance of less than 0.77 arcminutes from the analysis of MS stars in order to avoid the most crowded inner area.
 The stellar positions derived from the ACS/WFC and UVIS/WFC3 images are corrected for distortion by using the solutions provided by \citet{anderson2006a},\citet{bellini2009a}, and \citet{bellini2011a}. The photometry has been calibrated into the Vega-mag system by using the procedure outlined by \citet{milone2023b} and by adopting the encircled energy and the zero points available at the STScI web page\footnote{https://www.stsci.edu/hst/instrumentation/acs/data-analysis/zeropoints; \\ https://www.stsci.edu/hst/instrumentation/wfc3/data-analysis/photometric-calibration; \\ https:\,//jwst-docs.stsci.edu\,/jwst-near-infrared-camera\,/nircam-performance\,/nircam-absolute-flux-calibration-and-zeropoints}.

We exploited stellar proper motions to separate field stars from cluster members. We calculate the proper motions of stars in the central field as in \cite{milone2023b}, by comparing the stellar positions of images collected at different epochs. For the external field, we used the proper motions by \cite{marino2024a}.

\begin{table*}[h]
\centering
\begin{tabular}{ccccccc}

\toprule
 Filter & Instrument & N $\times$ Exp. time & date & Program  \\ 
\midrule
       &           &  Central field  & & \\
\hline
 F275W & UVIS/WFC3 & 2$\times$323s$+$12$\times$348s & Nov 21 2011   & 12311  \\ 
 F336W & UVIS/WFC3 & 30s$+$2$\times$580s & Sep 28-29 2010   & 11729  \\ 
 F343N & UVIS/WFC3 & 730s$+$1325s & Mar 23 2018   & 15061  \\ 
 F435W & ACS/WFC   & 10s$+$6$\times$100s$+$3$\times$115s & Sep 30 - Oct 11 2002 & 9281 \\
 F438W & UVIS/WFC3 & 40s$+$130s$+$435s & Mar 23 2018   & 15061  \\ 
 F606W & ACS/WFC   &  3s$+$4$\times$50s & Mar 13 2006 & 10775 \\
 F814W & ACS/WFC   &  3s$+$4$\times$50s & Mar 13 2006 & 10775 \\
 \hline
        &           &  External Field  & & \\
        \hline
 F336W & UVIS/WFC3 & 2$\times$300s & Sep 4 2015 - Mar 11 2016 & 14021 \\
 F435W & ACS/WFC   & 36$\times$30s & Apr 17 2002 - May 06 2006 & 9018 \\
 F435W & ACS/WFC   & 12$\times$350s & Sep 2 2006 & 10730 \\
 F435W & ACS/WFC   &  6$\times$339s & Nov 27-28 2005 & 10730 \\
 F814W & ACS/WFC   &  2$\times$1390s$+$2$\times$1460s & Oct 9 2002 & 9444 \\
 F606W & ACS/WFC   & 4$\times$1s$+$4$\times$10s$+$4$\times$100s$+$117$\times$1113s-1498s & Jan 15 2010 - Oct 1 2010 & 11677 \\
 F814W & ACS/WFC   & 4$\times$1s$+$4$\times$10s$+$4$\times$100s$+$125$\times$1031s-1484s & Jan 15 2010 - Oct 1 2010 & 11677 \\
 F160W & IR/WFC3   & 24$\times$274s & Jul 23 2009 & 11445 \\
 F160W & IR/WFC3   & 18$\times$274s & Jul 23 2009 & 11453 \\
 F160W & IR/WFC3   & 24$\times$92s$+$24$\times$352s & Mar 3 2010 - Nov 20 2010 & 11931 \\
 F160W & IR/WFC3   & 14$\times$92s$+$6$\times$352s & Apr 9 2011 - Sep 18 2011 & 12352 \\
 F160W & IR/WFC3   & 14$\times$92s$+$6$\times$352s & Feb 14 2012 - Aug 22 2012 & 12696 \\
 F160W & IR/WFC3   & 5$\times$92s$+$2$\times$352s & Apr 19 2013 & 13079 \\
 F160W & IR/WFC3   & 5$\times$92s$+$2$\times$352s & Dec 21 2013 & 13563 \\
 F322W2 & NIRCam & 48$\times$857s & Sep 14-15 2022 & 2560 \\
\bottomrule
\end{tabular}
\caption{Summary of the properties of the {\it HST} and {\it JWST} used in this work.}
\label{tab:data}
\end{table*}

\subsection{Artificial star tests}
To estimate the photometric errors and create the simulated color-magnitude diagram (CMD), we conducted artificial star (AS) tests for each field. Following the method outlined by \cite{anderson2008a}, we generated a list of 10$^6$ ASs with radial distributions and luminosity functions matching those of the observed stars. We adopted for the ASs the F814W instrumental magnitudes that range from the saturation limit to $-$5.0 mag\footnote{Instrumental magnitudes are defined as the $-$2.5 log$_{10}$ of the detected photo-electrons.}. The photometry and astrometry of the ASs and the selection of the relatively isolated ASs that are well fitted by the PSF model have been carried out with the same procedure used for real stars.

\section{Binaries among multiple populations }
\label{sec:binaries}
 In this section, we describe the procedure that we used to derive the fraction of binaries among 1P and 2P stars of 47\,Tucanae and present the results.
 In a nutshell, this method, detailed in the following subsections, is based on \citet{milone2020a} and \citet{muratore2024a} who estimated the binary incidence among multiple populations in Galactic GCs and Magellanic Cloud young star clusters. It utilizes two photometric diagrams: a CMD, where 1P and 2P stars appear indistinguishable and are used to identify binaries, and another diagram where 1P and 2P stars form distinct sequences. Binary fractions for 1P and 2P stars are inferred by comparing the color distributions of observed binaries with those from simulated CMDs containing varying 1P and 2P binary fractions.
 We separately analyzed stars with different luminosities and radial distances from the cluster center. Section\,\ref{subsec:mdwarfs} is focused on binaries among M-dwarfs 
   in the external field\footnote{ While NIRCam data are unavailable for the cluster center, the {\it HST} archive includes IR/WFC images in F110W and F160W (GO-11664, PI.,Brown). These filters can identify multiple populations below the MS knee, but crowding makes distinguishing 1P and 2P stars in the cluster center challenging. Hence, we only investigate binaries among M-dwarfs in the external regions.},  whereas Section\,\ref{subsec:ms} is dedicated to the investigation of upper-MS stars 
   in the central field.

\subsection{The M-dwarfs}\label{subsec:mdwarfs}
To estimate the fraction of binaries among the multiple populations of M-dwarfs in 47\,Tucanae, we used the method illustrated in Figures\,\ref{fig:selbinarie}--\ref{fig:simu}, 
 which is based on the procedure introduced by \citet{milone2020a} to derive the fractions of binaries composed of 1P and 2P stars along the upper MS of GCs.
 
\begin{figure*}[htp]
    \centering
    \includegraphics[width=.75\textwidth,trim={1cm 5.5cm 0.0cm 8cm},clip]{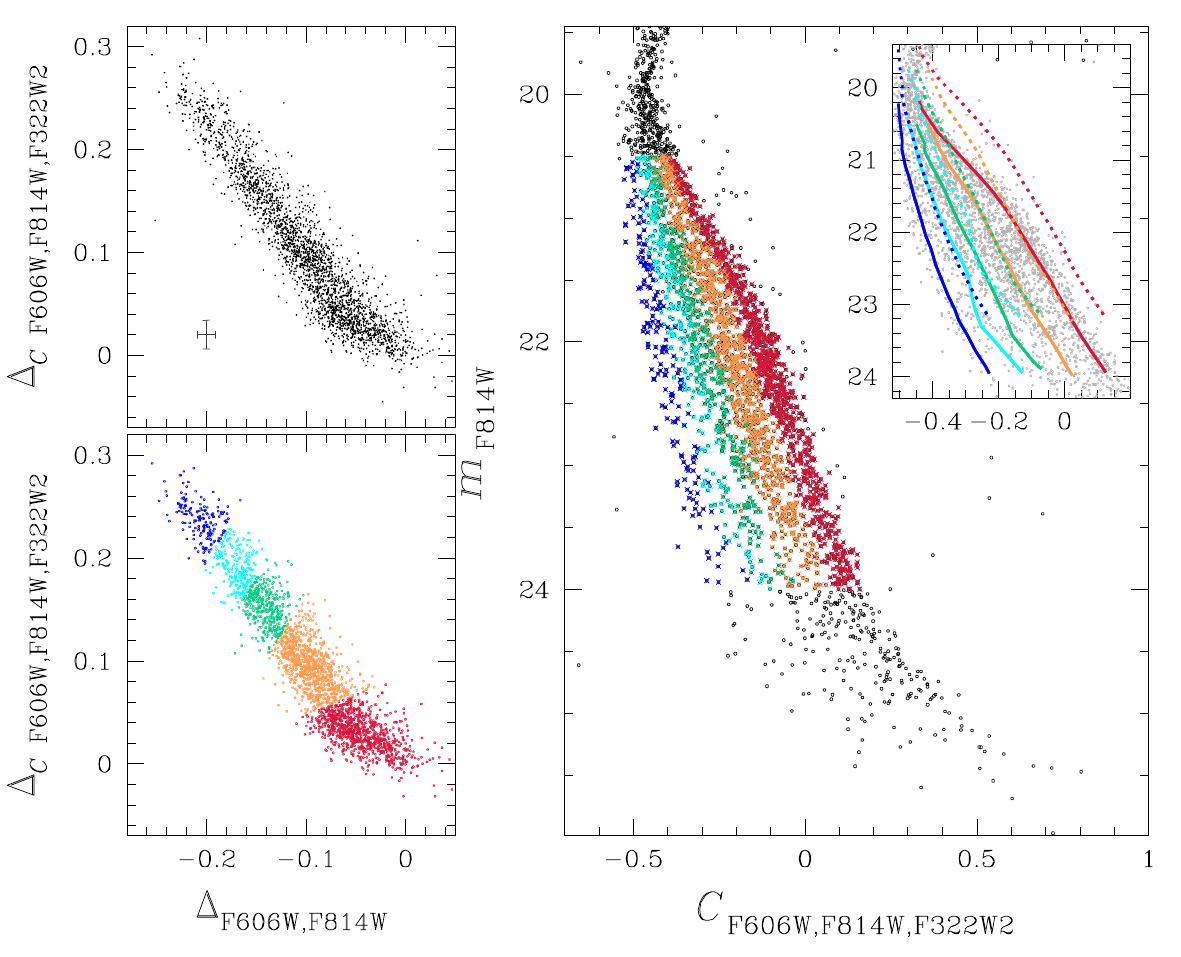}
    \caption{ $\Delta_{C {\rm F606W,F814W,F322W2}}$ vs.\,$\Delta_{\rm F606W,F814W}$ ChMs for M-dwarfs with $20.5<m_{\rm F814W}<24.00$ mag (left panels) and $m_{\rm F814W}$ vs.\,$C_{\rm F606W,F814W,F322W2}$ pseudo-CMD of M-dwarfs from \citet{marino2024a} (right panel). The probable 1P, 2P$_{\rm A}$, 2P$_{\rm B}$, 2P$_{\rm C}$, and 2P$_{\rm D}$ stars, selected from the ChM, are colored crimson, orange, green, cyan, and blue, respectively. The inset in the right panel is a zoom of the $m_{\rm F814W}$ vs.\,$C_{\rm F606W,F814W,F322W2}$ pseudo-CMD around the MS region where the five stellar populations are more-clearly distinguishable. The fiducial lines of the five stellar populations are represented with continuous colored lines, while the dotted lines represent the corresponding fiducial lines for equal-luminosity binaries. } 
    \label{fig:chm}
\end{figure*}

The first step consists in selecting groups of bona-fide 1P and 2P stars. To do this, we used the $\Delta_{C {\rm F606W, F814W, F322W2}}$ 
vs.\,$\Delta_{\rm F606W,F814W}$ ChM of M-dwarfs introduced by \citet{marino2024a}, which is reproduced in the top-left panel of Figure\,\ref{fig:chm}. This ChM
 is composed of MS stars with $20.0<m_{\rm F814W}<24.0$ mag, thus including the MS region where the sequences of 1P and 2P stars are more evident.
 As discussed by Marino and collaborators, 1P stars populate the ChM region near the origin of the reference frame, whereas the 2P defines a narrow sequence 
 of stars that extend towards the top-left corner of the ChM. The distribution of stars along the 2P sequence is not uniform but shows stellar overdensities that correspond to sub-populations of stars with different oxygen abundances \citep[][]{marino2024a, marino2024b}. 
 We identified a sample of bona-fide 1P stars, which are colored crimson in the ChM plotted in the bottom-left panel of Figure\,\ref{fig:chm}, and four main 
 groups of 2P stars, which we indicated as 2P$_{\rm A}$, 2P$_{\rm B}$, 2P$_{\rm C}$, and 2P$_{\rm D}$, and colored orange, aqua, cyan, and blue, respectively.

\begin{figure*}
    \centering
    \includegraphics[width=.75\textwidth,trim={1cm 5.0cm 0.0cm 11cm},clip]{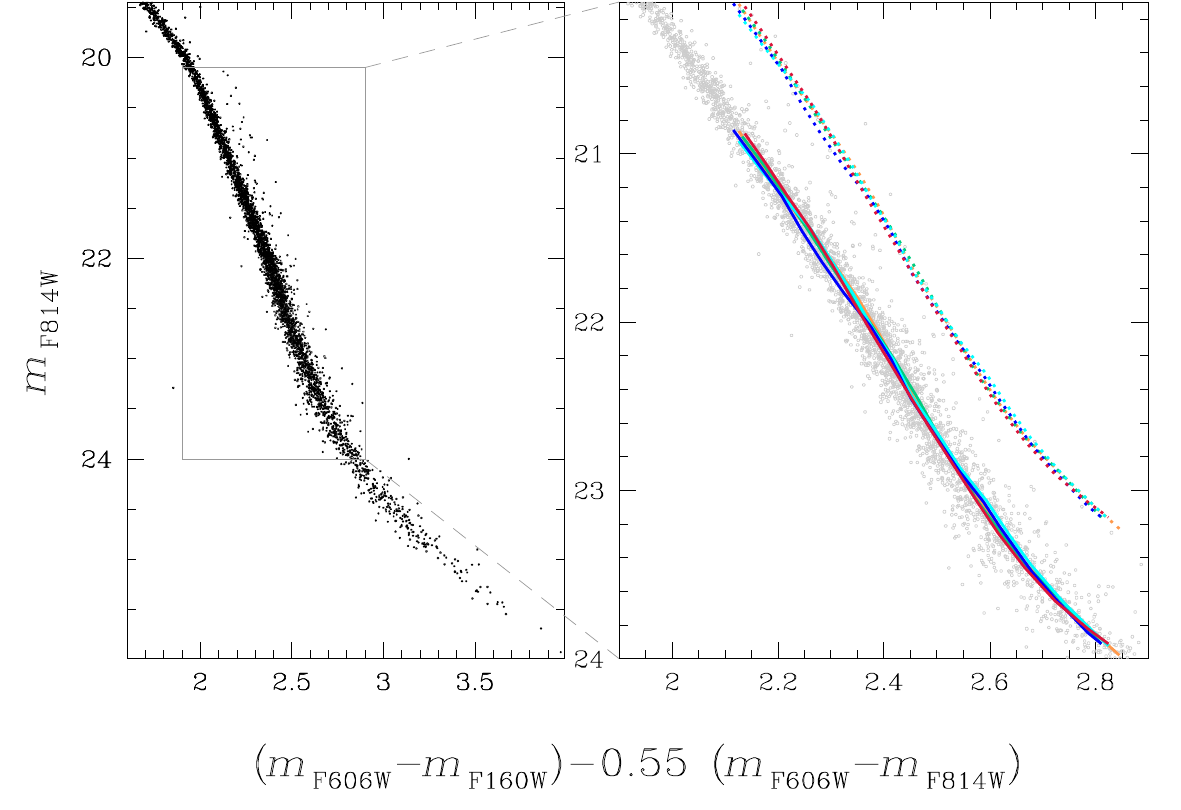}
    \caption{ \textit{Left.} $m_{\rm F814W}$ as a function of the color difference ($m_{\rm F606W}-m_{\rm F160W}$)$-$0.55($m_{\rm F606W}-m_{\rm F814W}$) for the stars plotted in Figure\,\ref{fig:chm}. \textit{Right.} Zoom in of the pseudo-CMD plotted in the left panel around the MS region used to derive the ChM. The continuous colored lines, which are partially overlapped with each other, are the fiducial lines of the five populations identified in the ChM of Figure\,\ref{fig:chm}. The fiducials of the corresponding sequences of equal-luminosity binaries are represented with dotted lines. } 
    \label{fig:Fiducials}
\end{figure*}

\begin{figure*}[htp!]
    \centering
    \includegraphics[height=.425\textwidth,trim={1cm 6.0cm 0.5cm 9cm},clip]{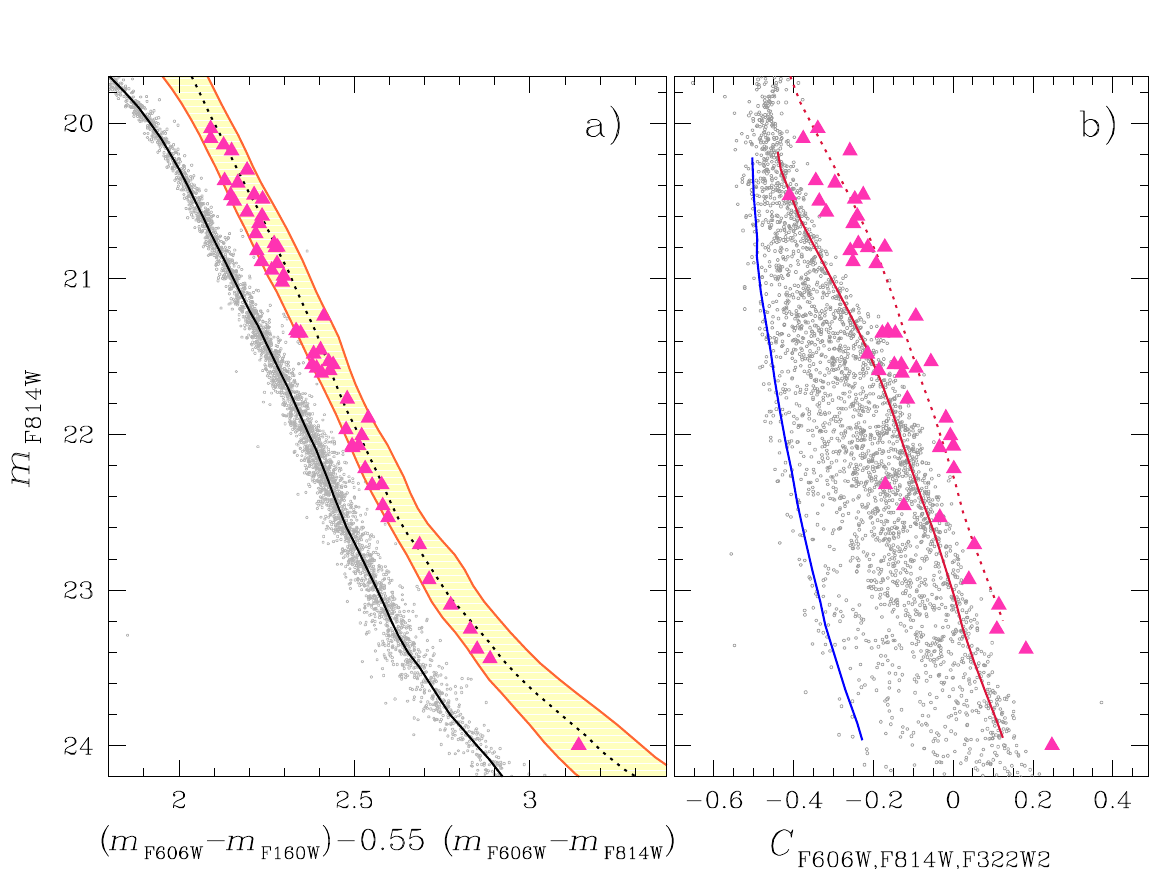}    
    \includegraphics[height=.425\textwidth,trim={3.1cm 6.0cm 4.5cm 9cm},clip]{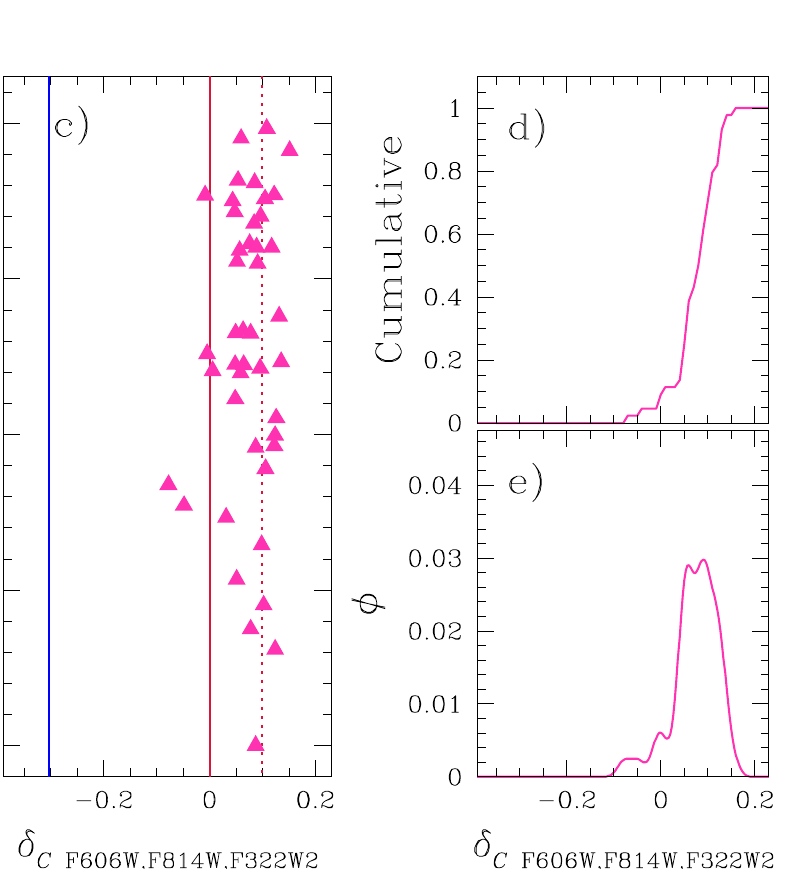}
    \caption{ Reproduction of the  $m_{\rm F814W}$ vs.\,($m_{\rm F606W}-m_{\rm F160W}$)$-0.55$($m_{\rm F606W}-m_{\rm F814W}$) pseudo-CMD of Figure\,\ref{fig:Fiducials} (panel a). The MS fiducial is represented with the continuous black line, whereas the dotted line is the fiducial of equal-luminosity binaries.  The orange lines mark the boundaries of the yellow shaded region that we used to select the binaries (magenta triangles). Panel b shows the $m_{\rm F814W}$ vs.\,$C_{\rm F606W, F814W, F322W2}$ pseudo-CMD. The fiducials of 1P and 2P$_{\rm D}$ stars are represented  with continuous red and blue lines, respectively, and the equal-luminosity 1P binaries fiducial is indicated by the red dotted line.  The verticalized $m_{\rm F814W}$ vs. $\delta_{C \rm {F606W, F814W, F322W2}}$ diagram is plotted in panel c, while the $\delta_{C \rm {F606W, F814W, F322W2}}$ cumulative and kernel-density distributions are plotted in panels d and e, respectively.
    }
    \label{fig:selbinarie}
\end{figure*}

\begin{figure*}[htp!]
    \centering
    \includegraphics[height=.425\textwidth,trim={1cm 6.0cm 0.5cm 9cm},clip]{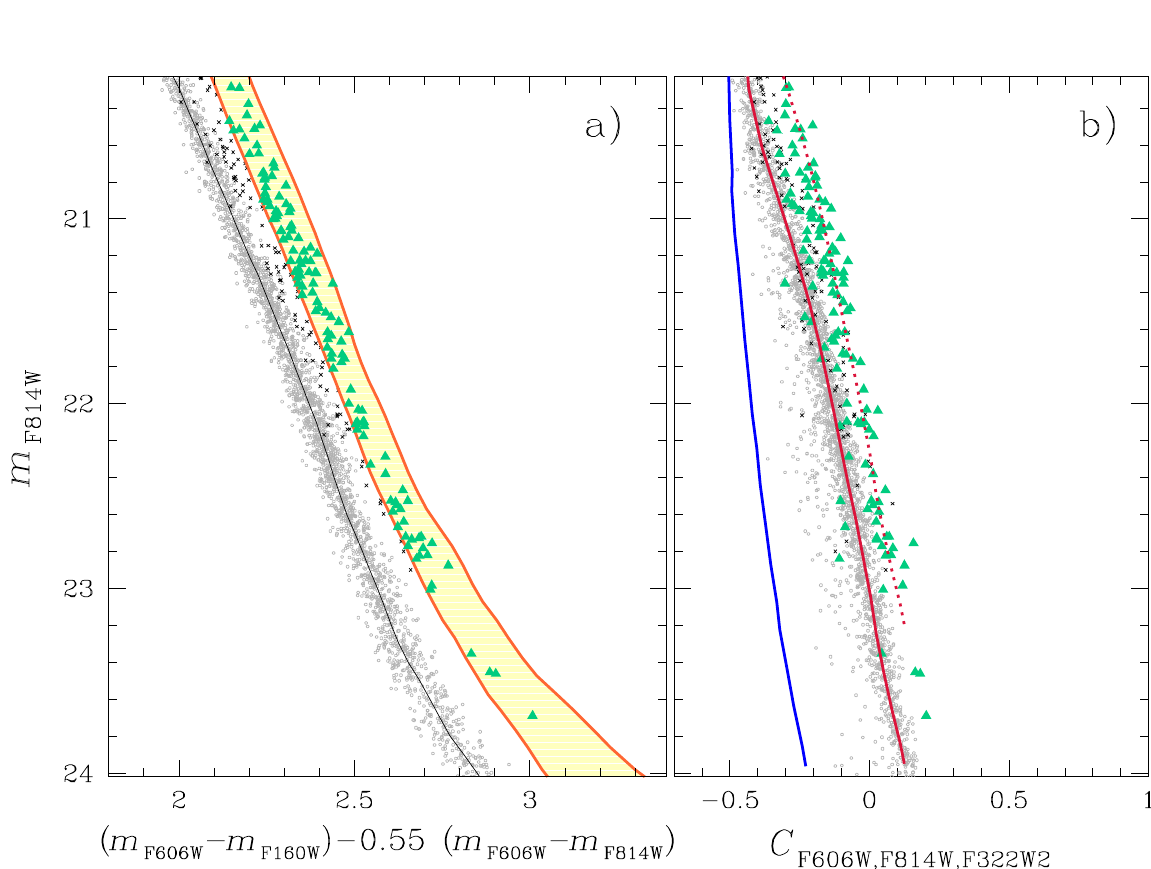}
    \includegraphics[height=.425\textwidth,trim={9.1cm 6.0cm 4.5cm 9cm},clip]{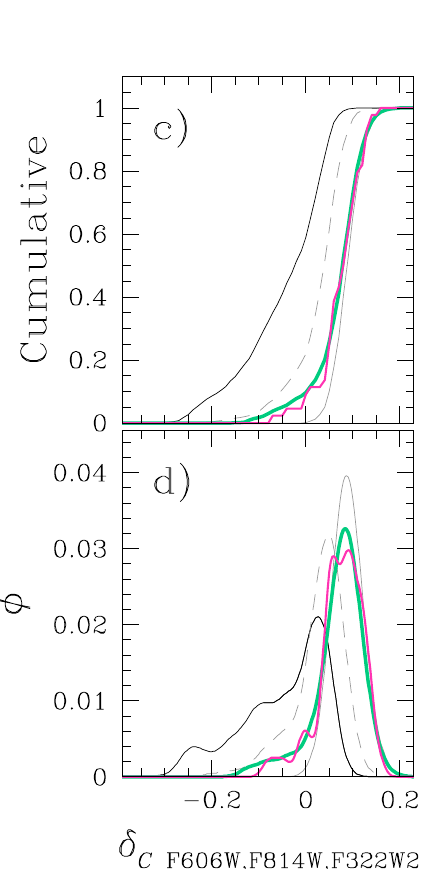}
    \caption{Simulated   $m_{\rm F814W}$ vs.\,($m_{\rm F606W}-m_{\rm F160W}$)$-0.55$($m_{\rm F606W}-m_{\rm F814W}$) (panel a) and $m_{\rm F814W}$ vs.\,$C_{\rm F606W, F814W, F322W2}$ (panel b) pseudo-CMDs. The adopted fractions of 1P, 2P, and mixed binaries provide the best fit with the observations. The selected binaries, which are located within the yellow shaded area of the panel-a diagram, are marked with aqua triangles, whereas the remaining binaries are represented with black symbols. For clarity, the number of plotted simulated binaries is three times larger than the number of observed binaries. The lines plotted in  panels a and b reproduce the fiducial lines that we defined in Figure\,\ref{fig:selbinarie}. Panels c and d compare the $\delta_{C \rm {F606W, F814W, F322W2}}$ cumulative and the kernel density distributions, respectively, for the observed binaries (magenta lines) and the best-fit simulations (aqua lines). For completeness, we show the distributions of 1P binaries (gray continuous lines), 2P binaries (black continuous lines), and mixed binaries (gray dashed lines).
    }
    \label{fig:simu}
\end{figure*}

Then, we used two pseudo-CMDs, where the multiple populations exhibit very different behaviours. 
  In the $m_{\rm F814W}$ vs.\,$C_{\rm F606W, F814W, F322W2}$ diagram plotted in the right panel of Figure\,\ref{fig:chm} the five stellar populations 
 identified in the ChM define distinct sequences. The pseudo-color separation among the sequences is well illustrated in the inset, where we superimposed 
 their fiducials (continuous lines) and the corresponding fiducials of equal-luminosity binaries\footnote{The fiducial line for binary systems, consisting of pairs of stars with identical magnitudes, aligns with the fiducial line of the corresponding single stars but is shifted brighter by $\sim$0.753 magnitudes.} (dotted lines) on the pseudo-CMD.
  Furthermore, we introduced the $m_{\rm F814W}$ vs.\,($m_{\rm F606W}-m_{\rm F160W}$)$-$0.55($m_{\rm F606W}-m_{\rm F814W}$) pseudo-CMD plotted in 
  Figure\,\ref{fig:Fiducials}, where the fiducial lines of the five selected stellar groups are nearly superimposed on each other.

The latter diagram is used to select a sample of binary systems composed of stars with similar luminosities as illustrated in Figure\,\ref{fig:selbinarie}a. 
 The 44 selected binaries are represented with large magenta triangles and are located within the yellow-shaded portion of the pseudo-CMD and delimited by the 
 two orange fiducial lines. Specifically, the left boundary is the fiducial line composed of MS-MS binaries where the secondary component is 1.6 mag fainter than the primary star in the F814W band. Based on diagrams constructed with ASs, this adopted value allows us to minimize the contamination of single stars with large observational errors.
 The right boundary is the fiducial line of equal-luminosity binaries shifted to the red by three times the pseudo-color 
 observational error.
 Figure\,\ref{fig:selbinarie}b highlights the selected binaries in the $m_{\rm F814W}$ vs.\,$C_{\rm F606W, F814W, F322W2}$ pseudo-CMD together with the fiducial
 lines of 1P and 2P$_{\rm D}$ stars (red and blue continuous lines) and the fiducial line of equal-luminosity 1P binaries (red-dotted line).
 These three fiducials are used to derive the verticalized $m_{\rm F814W}$ vs.\,$\delta_{ C {\rm F606W, F814W, F322W2}}$ plotted in panel c of Figure\,\ref{fig:selbinarie}, 
 which is obtained by using the procedure by \citet[][see their Appendix A]{milone2017a}.
 
In a nutshell, we considered two groups of stars in the $m_{\rm F814W}$ vs.\,$C_{\rm F606W, F814W, F322W2}$ pseudo-CMD, namely group I and II, which are located on the left and right side of the 1P fiducial line, respectively.
 We defined 
 \begin{align}  
     \delta_{ C {\rm F606W, F814W, F322W2}}=
     \begin{dcases*} \label{eq:1}
        W_{\rm I} \frac{X-X_{\rm fiducial\,A}}{X_{\rm fiducial\,B}-X_{\rm fiducial\,A}}, & \text{for group I},\\
        W_{\rm II} \frac{X-X_{\rm fiducial\,B}}{X_{\rm fiducial\,C}-X_{\rm fiducial\,B}}, & \text{for group II.}
        \end{dcases*}
  \end{align}
  \noindent
where $X= C_{\rm F606W,F814W,F322W2}$.
We have assumed for fiducial\,A, B, and C, the continuous blue, continuous red, and dotted-red fiducial lines of Figure\,\ref{fig:selbinarie}b, respectively.
 The constant $W_{\rm I}$ and $W_{\rm II}$ indicates the $\Delta_{C{\rm F606W,F814W,F322W2}}$ pseudo-color distance between the fiducials B and A at $m_{\rm F814W}=$22.25 mag. Similarly, the constant $W_{\rm II}$ refers to the distance between the fiducials C and B at the same F814W magnitude level.

 We derived the cumulative distribution of $\delta_{ C {\rm F606W, F814W, F322W2}}$ 
 and the corresponding kernel-density distribution, plotted in panels d and e of Figure\,\ref{fig:selbinarie}, respectively. The latter is obtained by assuming
 a Gaussian kernel with dispersion of 0.02 mag.

The fraction of binaries composed of  two 1P stars, two 2P stars, or a 1P and 2P star
(hereafter 1P, 2P, and mixed binaries) is derived by comparing the $\delta_{ C {\rm F606W, F814W, F322W2}}$ cumulative distribution of the observed binaries
 and the cumulative distributions derived from grids of simulated CMDs that are constructed by using the ASs. 
 We assumed grids of values that range from 0\% to 100\% in steps of 1\% for the fractions of 1P binaries (f$_{\rm bin}^{\rm 1P}$), 
  2P$_{\rm X}$ binaries (f$_{\rm bin}^{\rm 2PX}$, where X=A,B,C, and D) 
 and the fraction of mixed binaries (f$_{\rm bin}^{\rm MIX}$). For simplicity, we assumed that the 2P components of the mixed binaries have the same probability of
 belonging to each of the four populations 2P$_{\rm A--D}$, based on the relative fractions of stars in each of them calculated by \citet{marino2024a}.
For each combination of f$_{\rm bin}^{\rm 1P}$, f$_{\rm bin}^{\rm 2PX}$, and f$_{\rm bin}^{\rm MIX}$, we derived the $\delta_{ C {\rm F606W, F814W, F322W2}}$ cumulative distribution of the simulated binaries
 by using the same photometric diagrams and the same procedure adopted for the real stars. This quantity is compared with  the corresponding cumulative distribution
 of the observed binaries by means of $\chi^{2}$.

As summarized in Table\,\ref{tab:fbin}, we find a predominance of 1P binaries, with the fractions of 1P and 2P binaries corresponding to 84.8$\pm$5.0\%, and  8.2$\pm$4.5\% respectively. The inferred fraction of mixed binaries is 7.0$\pm$4.5\%.
The simulated diagrams that provide the best match with the observations are illustrated in the panels a and b of Figure\,\ref{fig:simu}, where we used aqua and black symbols to mark the simulated binary systems. The simulated cumulative and kernel-density distributions of the studied simulated binaries (aqua triangles in Figure\,\ref{fig:simu}a,b) are represented with aqua lines in the panels c and d, respectively, and are compared with the corresponding observed distributions that we colored magenta.

\begin{table*}[]
\centering
\begin{tabular}{lccccc}
\hline
Field & radius   & $m_{\rm F814W}$ interval & $f_{\rm bin}^{\rm 1P}$  & $f_{\rm bin}^{\rm MIX}$  & $f_{\rm bin}^{\rm 2P}$\\
& $[$arcmin] & [mag]                    &                         &                          & \\
\hline
External & 5.38-7.95 & 20.25-24.00 & 84.8$\pm5.0\%$ &  7.0$\pm4.5\%$ &  8.2$\pm$4.5\%   \\
Central  & 0.77-1.73 & 17.25-19.10 & 17.3$\pm7.3\%$ & 26.8$\pm7.3\%$ & 55.8$\pm$8.0\%   \\
Central  & 0.77-2.26 & 17.25-17.80 & 25.6$\pm9.5\%$ & 20.9$\pm8.3\%$ & 53.5$\pm$10.5\%   \\
External & 5.09-7.36 & 17.25-17.80 & 100$\pm6.5\%$ & 0.0$\pm6.5 \%$ & 0.0$\pm$6.5\%   \\
\hline
\end{tabular}
\caption{This table provides, for each dataset in the central and external fields, the studied radial and $m_{\rm F814W}$ magnitude intervals, and the fractions of 1P, mixed, and 2P binaries.}
\label{tab:fbin}
\end{table*}

To assess the uncertainties associated with the binary-fractions, we used ASs to create 10,000 pairs of mock $m_{\rm F814W}$ vs.\,($m_{\rm F606W}-m_{\rm F160W}$)$-$0.55 ($m_{\rm F606W}-m_{\rm F814W}$) and $m_{\rm F814W}$ vs.\,$C_{\rm F275W,F343N,F438W}$ photometric diagrams. These diagrams were generated with identical fractions of 1P, 2P and mixed binaries as determined from the observational data.
 We applied the identical methodology used for real stars to determine the fraction of binaries among the multiple stellar populations within each pair of simulated diagrams. For estimating the uncertainties associated with the observed fractions of 1P binaries, we assumed that the error corresponds to the root mean scatter of the 10,000 values of f$_{\rm bin}^{\rm 1P}$ derived from the simulated diagrams. We adopted a similar criterion to estimate the uncertainties associated with the best estimates of the fractions of 2P and mixed binaries.

\subsection{The upper main sequence}\label{subsec:ms}

\begin{figure*}
    \centering
    \includegraphics[width=.6\textwidth,trim={0.7cm 5.0cm 0.0cm 3cm},clip]{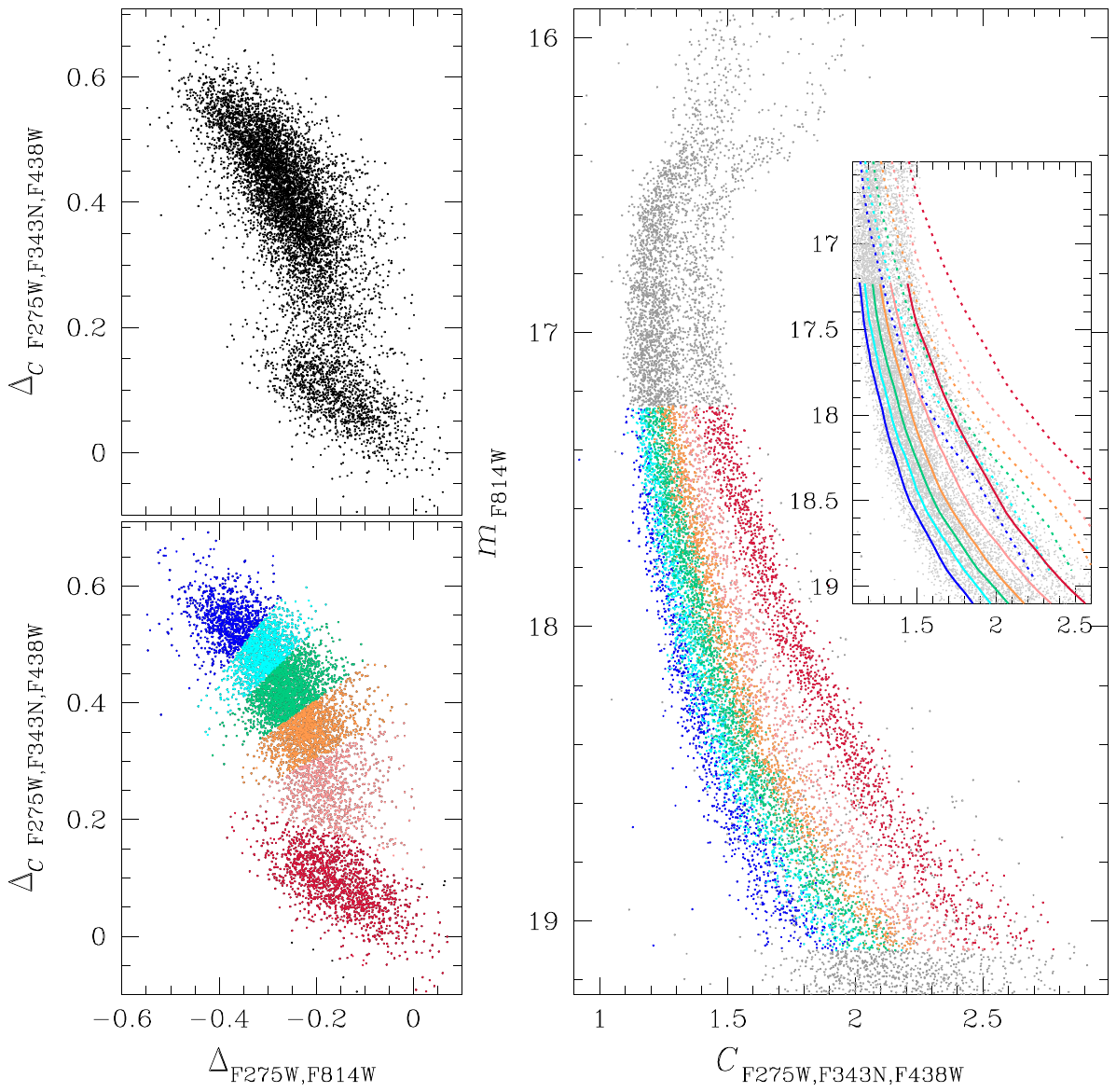}
    \caption{ $\Delta_{C \rm F275W,F343N,F438W}$ vs.\,$\Delta_{\rm F275W,F814W}$ ChMs for MS stars with $17.25<m_{\rm F814W}<19.10$ mag (left panels) and $m_{\rm F814W}$ vs.\,$C_{\rm F275W,F343N,F438W}$ pseudo-CMD (right panel) of stars in the central field from \cite{milone2022a}. The probable 1P stars are colored crimson, whereas the five groups of 2P stars, namely $\alpha$, $\beta$, $\gamma$, $\delta$, and $\epsilon$ that we selected from the ChM are colored pink, orange, aqua, cyan, and blue, respectively. The continuous lines superimposed on the diagram in the inset of the right panel represent the fiducial  lines of the six stellar populations, while the corresponding fiducial lines of equal-luminosity binaries are represented with dotted lines.}
    \label{fig:FiducialsHST}
\end{figure*}

To investigate the binaries among the bright MS stars ($17.25 <  m_{\rm F814W} < 19.10$ mag) we used the procedure described in Section\,\ref{subsec:mdwarfs} and by \cite{milone2020a}.
We took advantage of the  $\Delta_{C \rm {F275W,F343N,F438W}}$ vs.\,$\Delta_{\rm F275W,F814W}$ ChM derived by \cite{milone2022a} and reproduced in the left panels of Figure\,\ref{fig:FiducialsHST} to identify the probable 1P (crimson points in the bottom-left panel of Figure\,\ref{fig:FiducialsHST}) and 2P stars. In addition, we selected by eye four groups of 2P stars, namely 2P$_{\rm \alpha}$, 2P$_{\rm \beta}$, 2P$_{\rm \gamma}$, 2P$_{\rm \delta}$, and  2P$_{\rm \epsilon}$ that we colored pink, orange, aqua, cyan, and blue, respectively. 

We used the same colors to represent the selected stars in the $m_{\rm F814W}$ vs.\,$C_{\rm F275W,F343N,F438W}$ pseudo-CMD plotted in the right panel of Figure\,\ref{fig:FiducialsHST}. As shown in the inset, the selected stellar groups define distinct fiducial lines. 

\begin{figure*}[htp!]
    \centering
    \includegraphics[height=.425\textwidth,trim={1cm 6.0cm 0.5cm 9cm},clip]{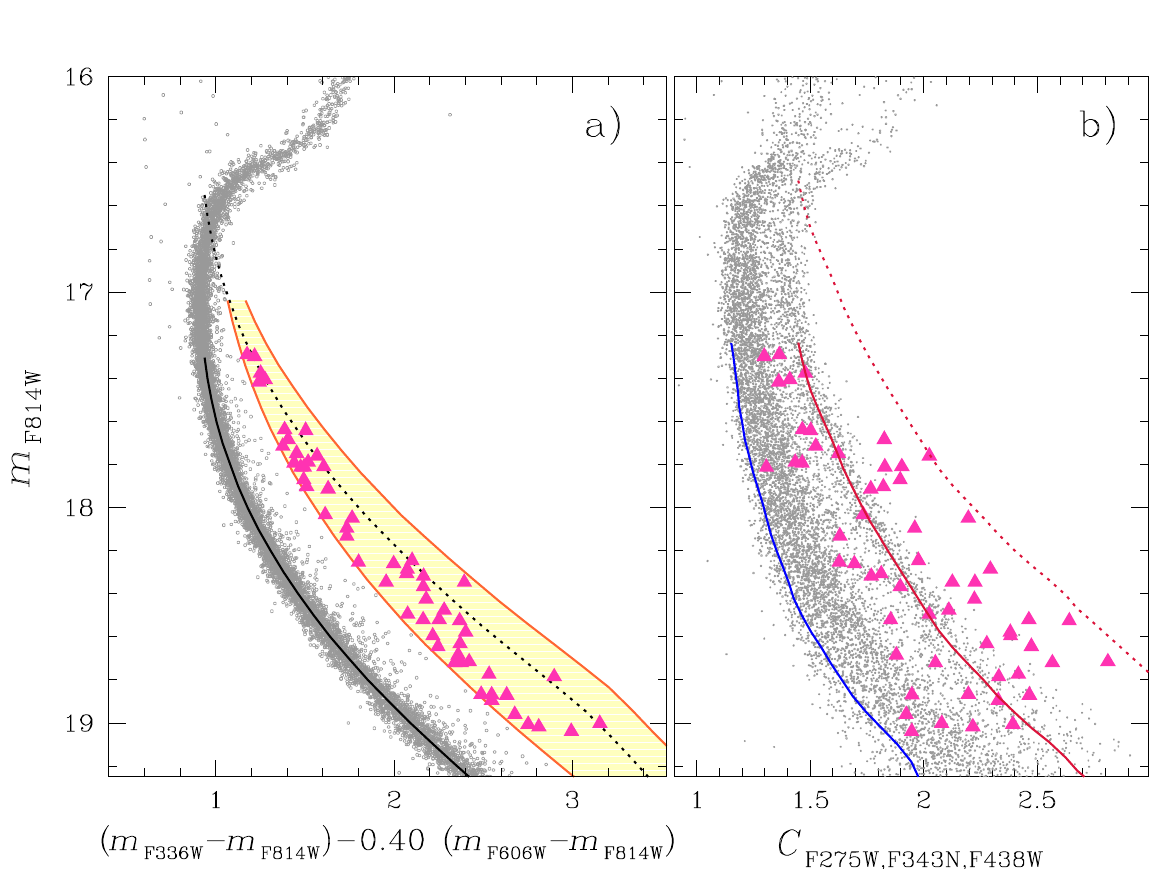}
    \includegraphics[height=.425\textwidth,trim={3.1cm 6.0cm 4.5cm 9cm},clip]{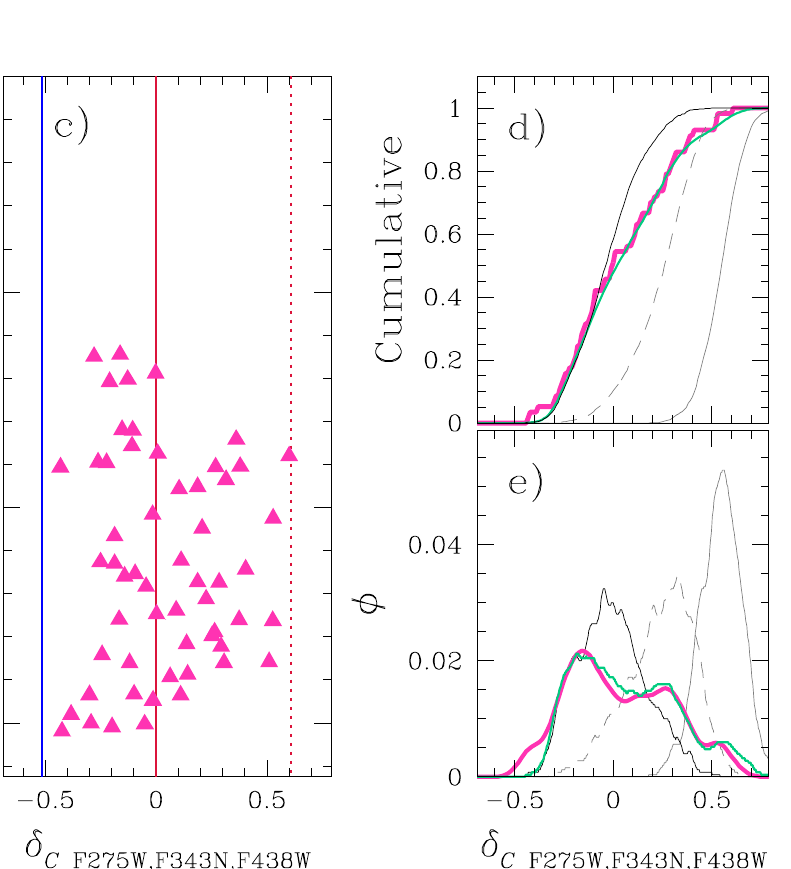}
    \caption{ $m_{\rm F814W}$ vs.\,($m_{\rm F336W}-m_{\rm F814W}$)$-$0.40\,($m_{\rm F606W}-m_{\rm F814W}$) (panel a) and $m_{\rm F814W}$ vs.\,$C_{\rm F275W,F343N,F438W}$ (panel b) pseudo-CMDs of stars in the central field. 
    Panel c shows the verticalized $m_{\rm F814W}$ vs.\,$\delta_{C {\rm F275W,F336W,F438W}}$ diagrams for the MS-MS binaries that we selected in panel a and represented with magenta triangles in panels a, b, and c. The $\delta_{C {\rm F275W,F336W,F435W}}$ cumulative and the kernel-density distributions of the studied binaries are illustrated with magenta lines in panels d and e, respectively, whereas the corresponding distributions for the simulated binaries that provide the best-fit with the observations are colored aqua. The gray and black lines correspond to the cumulative and the kernel-density distributions of 1P and 2P binaries alone, while the gray dashed lines represent the same distributions for mixed binaries.
    The continuous and the dotted black line plotted in panel a is the MS fiducial line and the fiducial of equal-luminosity binaries, whereas the orange lines delimit the yellow-shaded area of the CMD, which hosts the studied binaries. The blue lines plotted in panels b and c correspond to the fiducials of 2P$_{\rm \delta}$ stars, whereas the red-continuous and red-dotted lines are the fiducial lines of 1P stars and the fiducial lines of equal-luminosity 1P binaries, respectively. 
    }
    \label{fig:selbinarieHST}
\end{figure*}
On the contrary, we verified that the fiducials of 1P and 2P stars of 47\,Tucanae are nearly coincident in the $m_{\rm F814W}$ vs.\,($m_{\rm F336W}-m_{\rm F814W}$)$-$0.40($m_{\rm F606W}-m_{\rm F814W}$) pseudo-CMD (panel a of Figure\,\ref{fig:selbinarieHST}). Hence, we used this diagram to identify the sample of 57 probable binaries along the MS. The selected binaries are represented with magenta triangles and are located within the yellow-shaded region of the pseudo-CMD of Figure\,\ref{fig:selbinarieHST}a. The latter is defined as in Figure\,\ref{fig:selbinarie}a. In this case, the blue boundary corresponds to the fiducial of MS binaries with a F814W luminosity difference of 1.6 mag.

We used the fiducial lines of 2P$_{\rm \delta}$, 1P stars, and equal-luminosity 1P binaries superimposed on the $m_{\rm F814W}$ vs.\,$C_{\rm F275W,F336W,F438W}$ pseudo-CMD of Figure\,\ref{fig:selbinarieHST}b to derive the verticalized $m_{\rm F814W}$ vs.\,$\delta_{C {\rm F275W,F343N,F438W}}$ diagram plotted in Figure\,\ref{fig:selbinarieHST}c. To do that, we used the Equations\,\ref{eq:1} and assumed that W$_{\rm I}$ and W$_{\rm II}$ correspond to the $C_{\rm F275W,F336W,F438W}$ pseudo-color distance between the continuous blue and red fiducials and between the continuous red and dotted red fiducials, respectively, at $m_{\rm F814W}=18.3$ mag. The cumulative and the kernel-density distributions of the selected binaries are represented in panels d and e of Figure\,\ref{fig:selbinarieHST}.

To derive the fractions of 1P, 2P, and mixed binaries, we used the procedure outlined in Section\,\ref{subsec:mdwarfs} and compared the observed $\delta_{C {\rm F275W,F343N,F438W}}$ cumulative distributions and the cumulative distributions of grids of simulated binaries.
 The best match   is provided by a fraction of 17.3$\pm$7.3 \% 1P binaries, 55.8$\pm$8.0 \% 2P binaries, and 26.8$\pm$7.3 \% mixed binaries.

\subsection{The radial distribution of 1P and 2P binaries}\label{subsec:RD}
 The analysis presented in the previous subsections reveals a predominance of 1P binaries in the outer field, whereas the central field hosts significant amounts of 1P, 2P, and mixed binaries.
 However, the results in the central field are based on bright MS stars, whereas the study of the outer field is focused on M-dwarfs. 

\begin{figure*}[htp!]
    \centering
    \includegraphics[height=.425\textwidth,trim={1cm 6.0cm 0.5cm 9cm},clip]{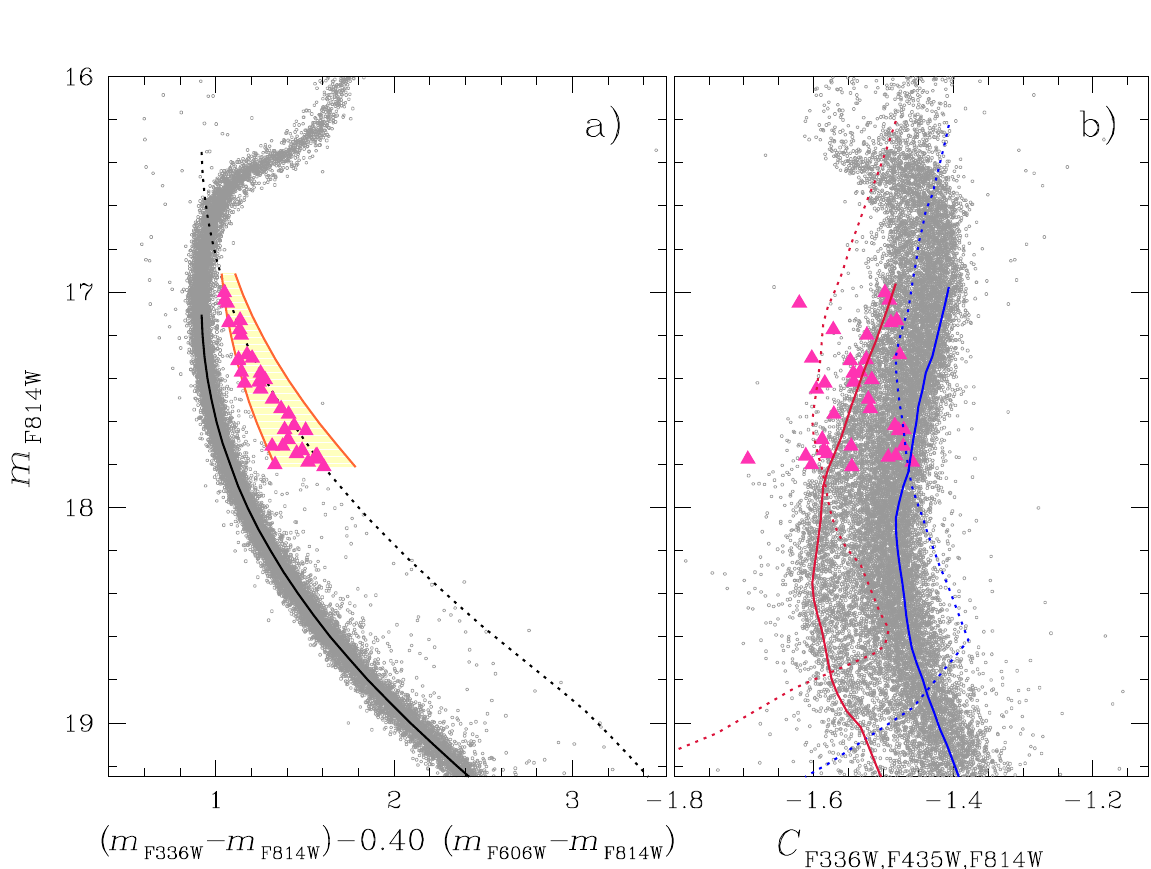}
    \includegraphics[height=.425\textwidth,trim={3.1cm 6.0cm 4.5cm 9cm},clip]{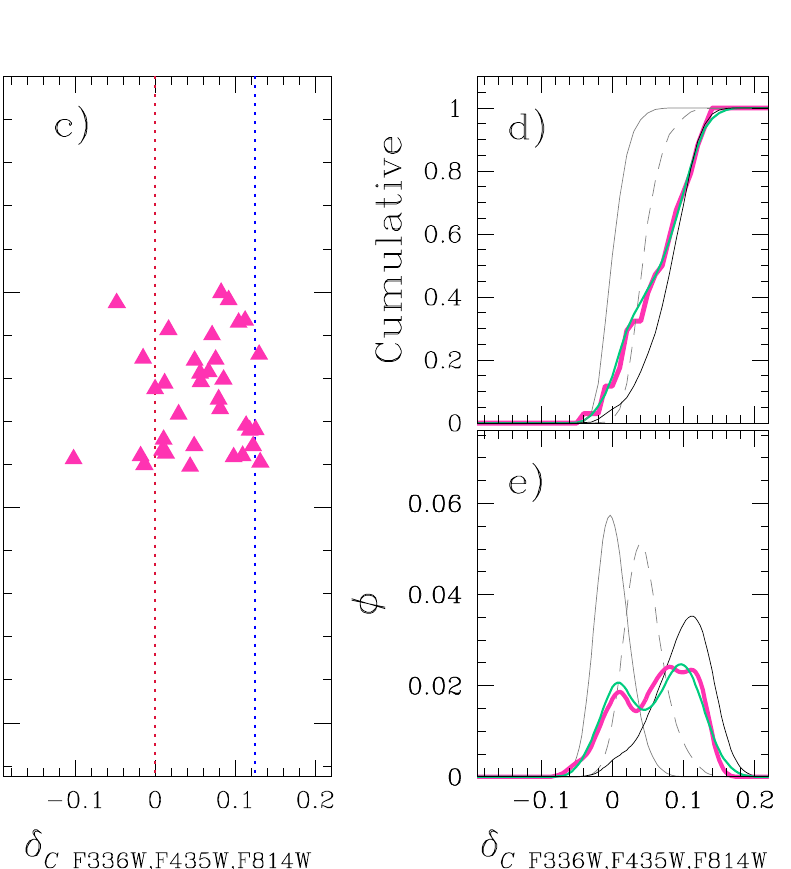}
    \caption{ Reproduction of the $m_{\rm F814W}$ vs.\,($m_{\rm F336W}-m_{\rm F814W}$)$-$0.40\,($m_{\rm F606W}-m_{\rm F814W}$) pseudo-CMD of stars in the central field plotted in Figure\,\ref{fig:FiducialsHST} (panel a). Panel b shows the 
     $m_{\rm F814W}$ vs.\,$C_{\rm F336W,F435W,F814W}$  pseudo-CMDs, whereas the  verticalized $m_{\rm F814W}$ vs.\,$\delta_{C {\rm F336W,F435W,F814W}}$ 
      diagram for the MS-MS binaries is plotted in panel c.
      The binaries are selected in panel a and represented with magenta triangles in panels a, b, and c. The red and blue lines superimposed on the diagram plotted in panel b are the fiducials of 1P and 2P$_\delta$ stars, respectively, while the corresponding fiducial lines of equal-luminosity binaries are represented with dotted lines  in panels b and c. Panels d and e show the cumulative and kernel-density $\delta_{C {\rm F336W,F435W,F814W}}$ distributions, respectively, for the observed binaries (magenta continuous lines), for the simulated binaries that provide the best match with the observations (aqua continuous lines), 1P binaries (gray continuous lines), mixed binaries (gray dashed lines), and 2P binaries (black continuous lines).
    }
    \label{fig:selbinarieCubi1}
\end{figure*}

\begin{figure*}[htp!]
    \centering
    \includegraphics[height=.425\textwidth,trim={1cm 6.0cm 0.5cm 9cm},clip]{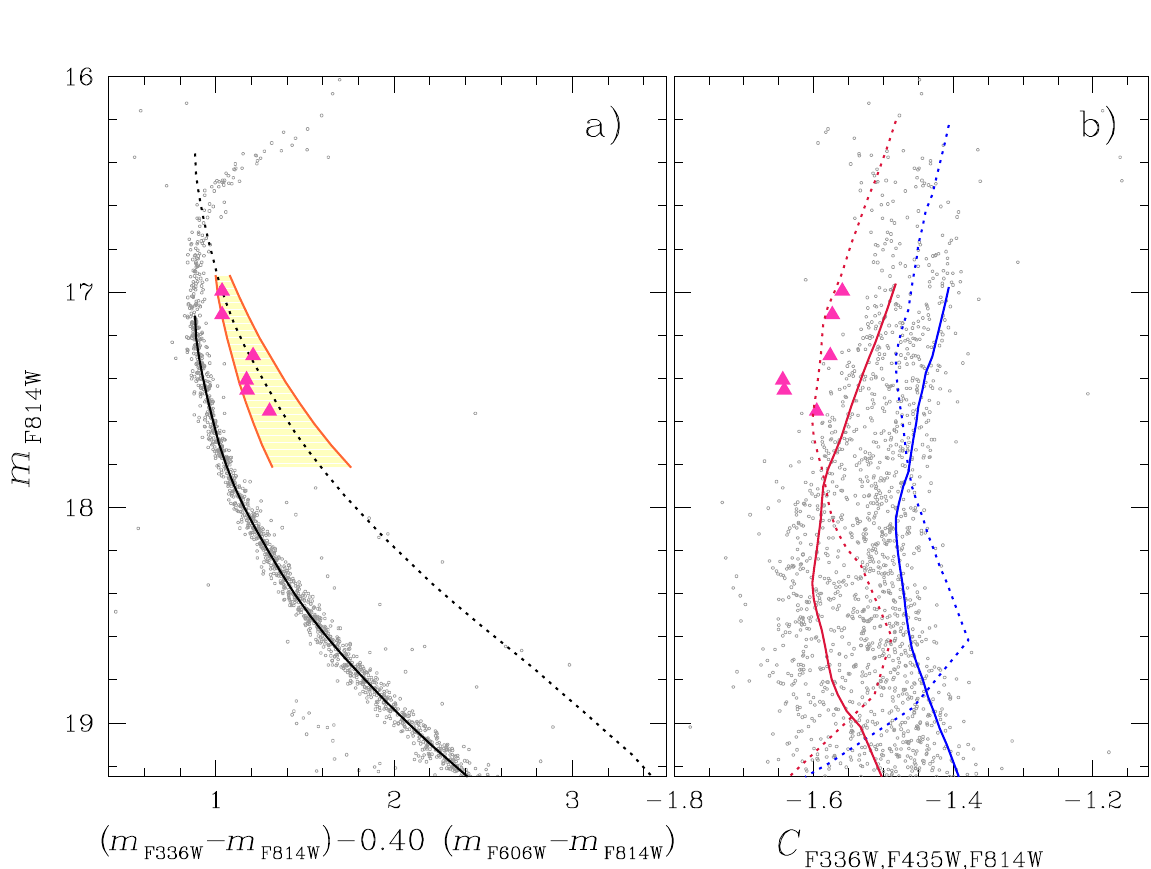}
    \includegraphics[height=.425\textwidth,trim={3.1cm 6.0cm 4.5cm 9cm},clip]{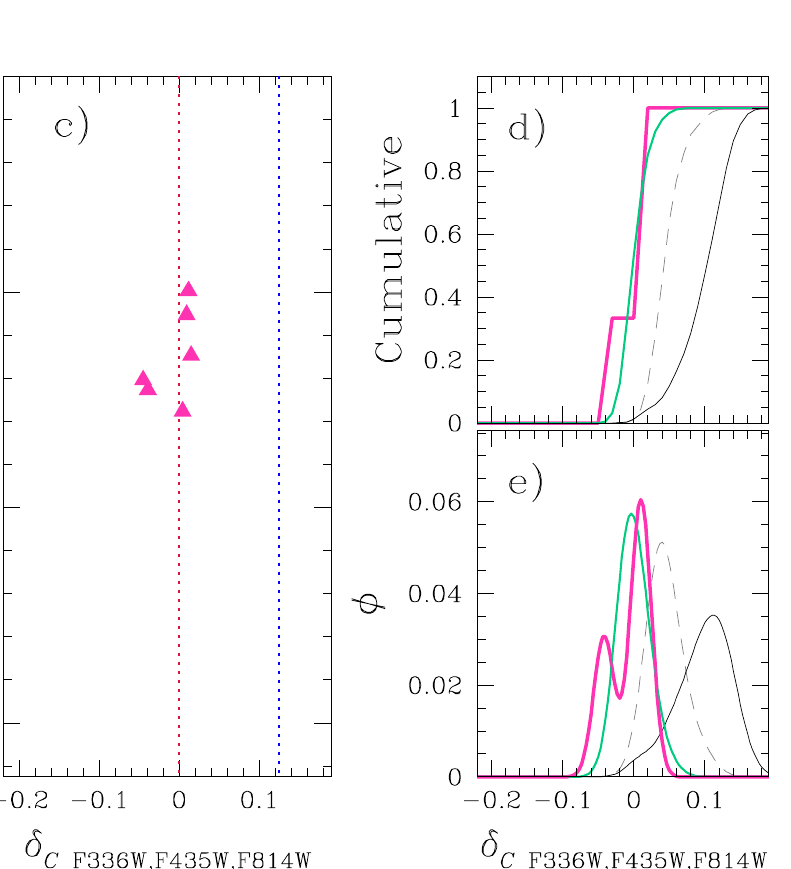}
    \caption{Same as in Figure\,\ref{fig:selbinarieCubi1} but for stars in the outer field.}
    \label{fig:selbinarieCubi2}
\end{figure*}

To ensure that the results on radial distribution are not influenced by stellar mass and to disentangle the effects of stellar mass and radial distance on the frequencies of binaries among multiple stellar populations, we used the photometry in the F336W, F435W, F606W, and F814W bands, which is available for bright MS stars in both the central and the outer field.
 The procedure that we used to study the binaries among 1P and 2P stars is similar to the one presented in Section\,\ref{subsec:ms} and is illustrated in the Figures\,\ref{fig:selbinarieCubi1} and \ref{fig:selbinarieCubi2} for stars in the central and external field, respectively.
Similarly to what we have done in Section\,\ref{subsec:ms}, we used the $m_{\rm F814W}$ vs.\,($m_{\rm F336W}-m_{\rm F814W}$)$-$0.40\,($m_{\rm F606W}-m_{\rm F814W}$) pseudo-CMD to identify the candidate binaries. In this case, we selected all candidate binary systems where the F814W luminosity of the two components differs by less than 1.8 mag (magenta triangles in Figures\,\ref{fig:selbinarieCubi1} and \ref{fig:selbinarieCubi2}).

The b panels of these figures highlight the selected binaries in the $m_{\rm F814W}$ vs.\,$C_{\rm F336W,F435W,F814W}$  pseudo-CMDs. These diagrams provide a smaller pseudo-color separation between 1P and 2P stars \citep[e.g.][]{jang2022a, milone2022a}. Moreover, the fiducial lines of equal-luminosity 1P and 2P binaries cross their respective single-stars fiducial lines around $m_{\rm F814W}=17.8$ and 19.0 mag, thus making it challenging to identify binaries among multiple populations for $m_{\rm F814W} \gtrsim 17.8$. Nevertheless, this diagram allows us to homogeneously analyse bright MS stars with different radial distances.

The c panels of Figures\,\ref{fig:selbinarieCubi1} and \ref{fig:selbinarieCubi2} show the $m_{\rm F814W}$ vs.\,$\delta_{C {\rm F336W,F435W,F814W}}$ 
      diagram for the selected binaries. In this case, we used the fiducial lines of equal-luminosity binaries to verticalize the pseudo-CMD. Finally, panels d and e of Figures\,\ref{fig:selbinarieCubi1} and \ref{fig:selbinarieCubi2} illustrate the $\delta_{C {\rm F336W,F435W,F814W}}$ cumulative and kernel-density distributions (magenta lines) of the selected binaries. We also show the distributions of the simulated binaries that provide the best fit with the observations (aqua lines), and of 1P binaries (gray continuous lines), mixed binaries (gray dashed lines), and 2P binaries (black lines).

We found that the central field hosts both 1P, 2P, and mixed binary systems, which comprise 25.6$\pm$9.5 \%, 53.5$\pm$10.5 \%, and 20.9$\pm$8.3 \% of the 35 studied  binaries, respectively. In contrast, the six binary stars that we studied in the outer-field are all composed of 1P pairs. The results of this section thus confirm the radial trend in the fraction of 1P and 2P binaries found in the previous subsection using M-dwarfs for the outer field.

\section{Chromosome map of binaries}\label{sec:chm}
In our opinion, the procedure introduced in the previous section is the optimal method for quantifying the fractions of binaries among 1P and 2P stars in GCs. Furthermore, the examination of the ChM can offer a qualitative yet direct perspective on the distribution of binaries among the different populations.

\begin{figure}[htp!]
    \centering
    \includegraphics[height=.325\textwidth,trim={0.5cm 5.0cm 0.5cm 11cm},clip]{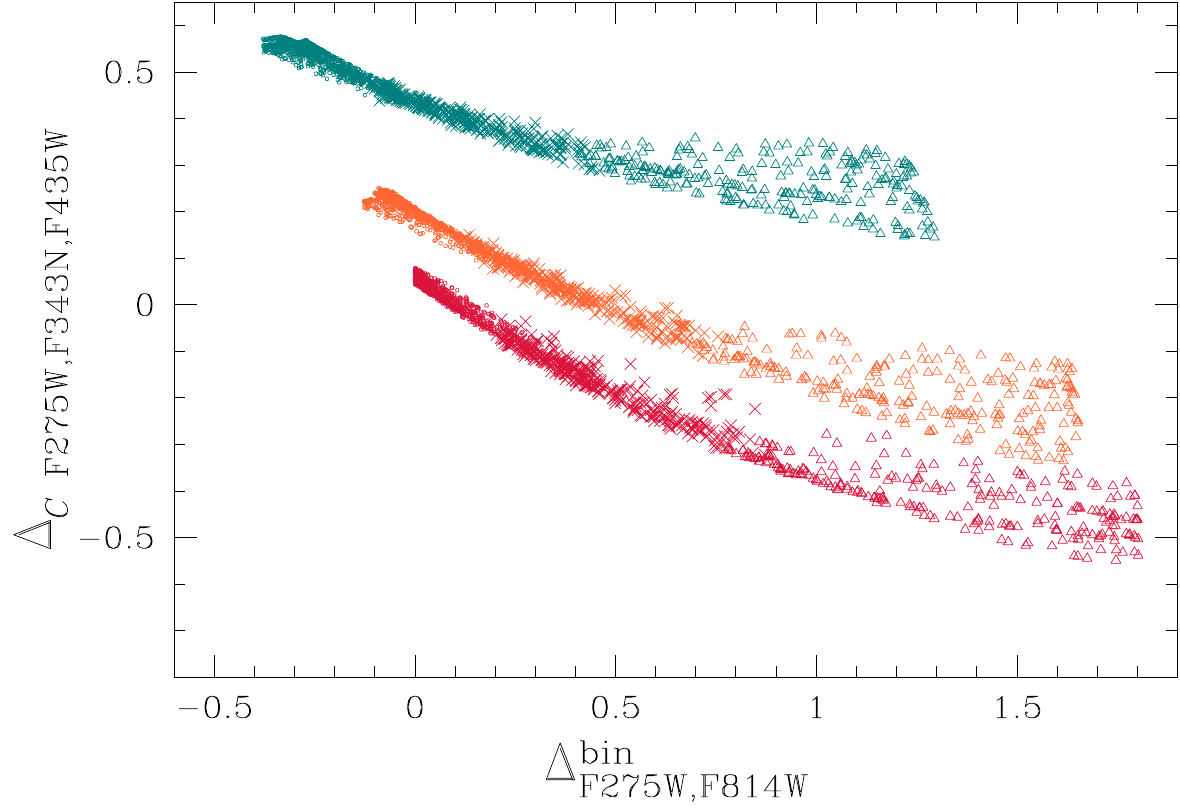}
    \includegraphics[height=.325\textwidth,trim={0.5cm 5.0cm 0.5cm 11cm},clip]{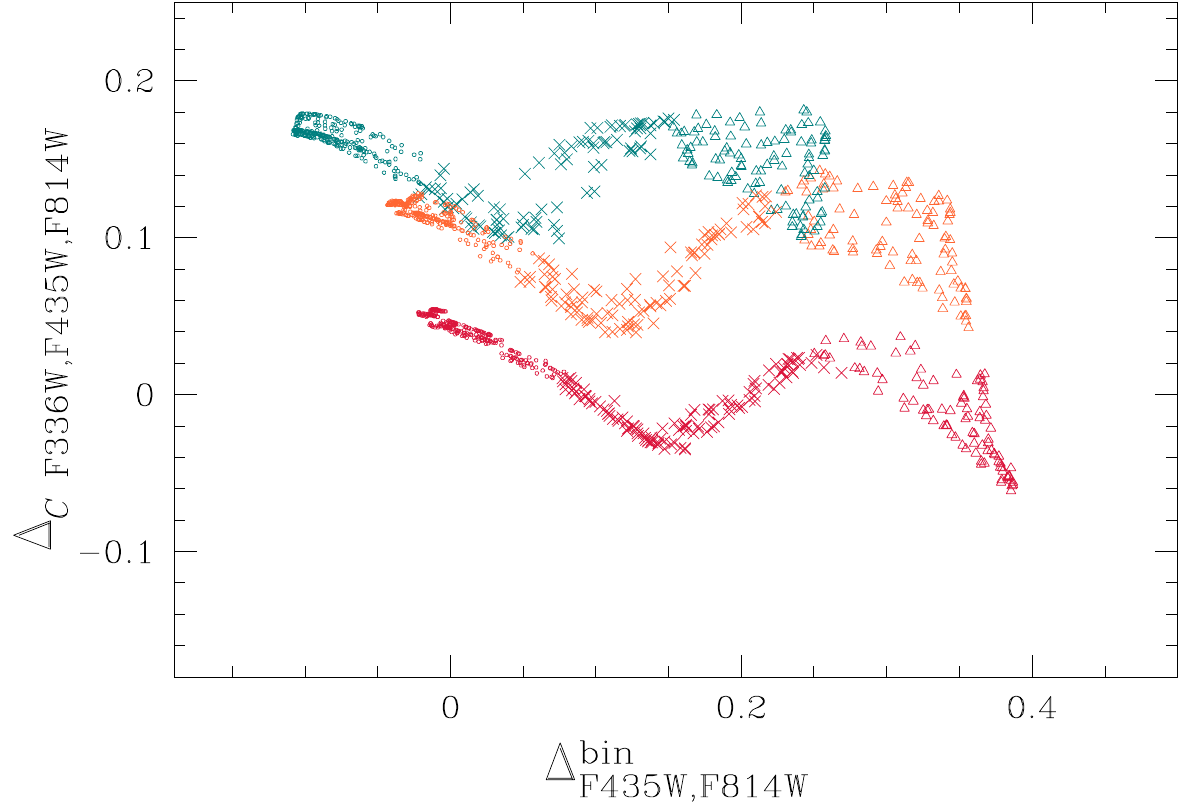}
            \includegraphics[height=.325\textwidth,trim={0.5cm 5.0cm 0.5cm 11cm},clip]{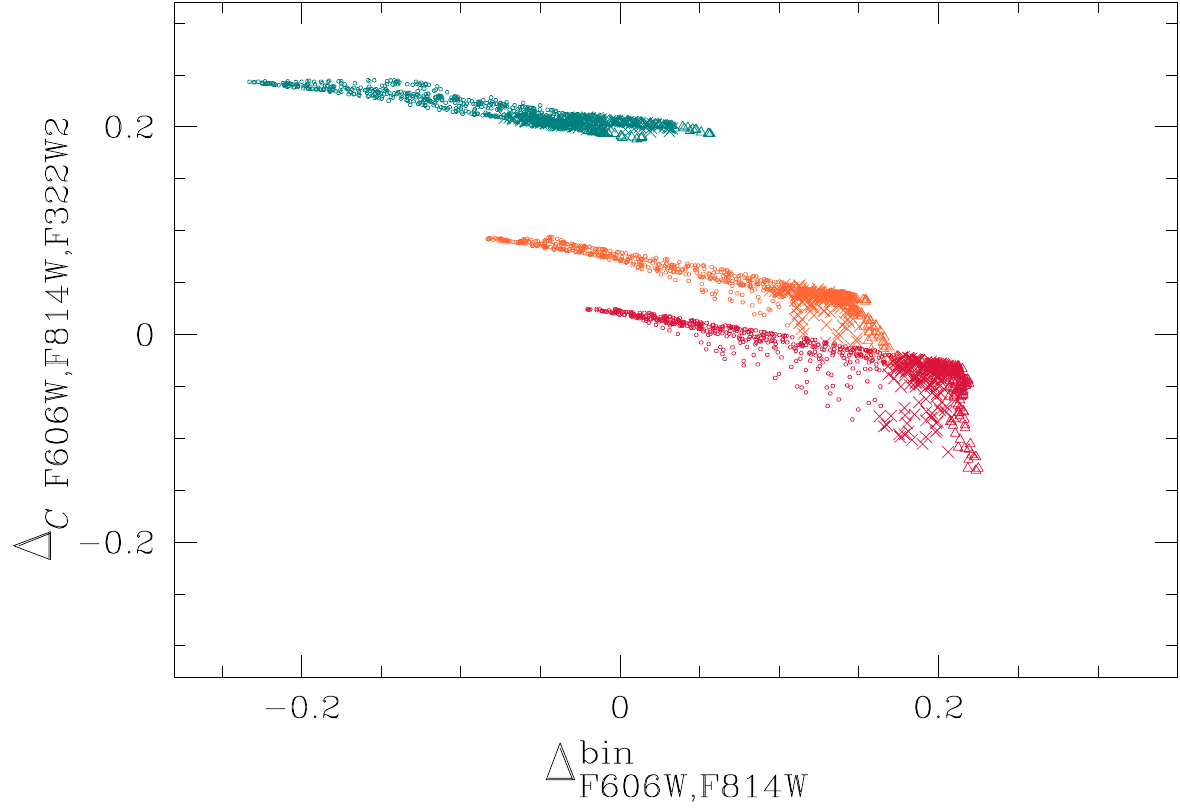}
    \caption{Simulated ChMs for upper-MS stars of 47\,Tucanae. Crimson, orange, and teal colors indicate 1P, 2P$_{\beta}$, and 2P$_{\epsilon}$ binaries, respectively. The triangles, crosses, and circles indicate the binaries with mass ratios q$>0.8$, $0.6<{\rm q}<0.8$, and q$<0.6$, respectively.}
    \label{fig:teochm}
\end{figure}

The top panel of Figure\,\ref{fig:teochm} shows the simulated $\Delta_{\rm {\it C} F275W,F343N,F435W}$ vs.\,$\Delta_{\rm F275W,F814W}^{\rm bin}$ ChM for MS binaries of 47\,Tucanae with $17.25<m_{\rm F814W}<19.10$ mag. 
 Specifically, the crimson, orange, and teal colors correspond to 1P, 2P$_{\beta}$, and 2P$_{\epsilon}$ binaries, respectively. 

The ordinate of this diagram is derived as in \cite{milone2017a} by using the $m_{\rm F814W}$ vs.\,$C_{\rm F275W,F343N,F438W}$ diagram. The abscissa is obtained from the $m_{\rm F814W}$  vs.\,$m_{\rm F275W}-m_{\rm F814W}$ CMD  and is calculated as:
 \begin{align}  
 \small
    \Delta_{\rm F275W,F814W}^{\rm bin}=
     \begin{dcases*} \label{eq:1}
        \Delta_{\rm F275W,F814W}, & \text{for $\Delta_{\rm F275W,F814W}\leq$0},\\
        W_{\rm bin} \Bigg{(} 1+\frac{X-X_{\rm fiducial\,bin}}{X_{\rm fiducial\,bin}-X_{\rm fiducial\,R}} \Bigg{)}, & \text{ for $\Delta_{\rm F275W,F814W}>$0}
        \end{dcases*}
  \end{align}
\noindent
   where the quantity $\Delta_{\rm F275W,F814W}$ is defined by \citet[][see their section 3.2]{milone2017a} and $W_{\rm bin}$ is the $m_{\rm F275W}-m_{\rm F814W}$ color separation at $m_{\rm F814W}=$18.3 mag between the fiducial line that marks the red boundary of the MS (fiducial\,R) and the corresponding fiducial line shifted by 0.752 mag to the bright side \citep[fiducial\,bin, see][for details]{milone2017a}.  
To simulate the binary systems, we combined the colors and the magnitudes of MS stars located on the fiducial lines of each population.
Stellar masses were inferred using the isochrones from \citet{dotter2008a}, which provide the best match with the $m_{\rm F814W}$ vs.\,$m_{\rm F606W}-m_{\rm F814W}$ CMD. For the binaries, we adopted a flat mass-ratio distribution.
We used the same method to generate $\Delta_{\rm {\it C} F336W,F435W,F814W}$ vs.\,$\Delta_{\rm F435W,F814W}^{\rm bin}$ and  $\Delta_{\rm {\it C} F606W,F814W,F322W2}$ vs.\,$\Delta_{\rm F606W,F814W}^{\rm bin}$ ChMs that we plotted in the middle and bottom panel of Figure\,\ref{fig:teochm} for MS binaries in the same magnitude intervals used for real stars. 
Clearly, the 1P and 2P binaries populate different regions of these ChMs.

\begin{figure*}[htp!]
    \centering
    \includegraphics[height=.325\textwidth,trim={0.5cm 5.0cm 0.5cm 11cm},clip]{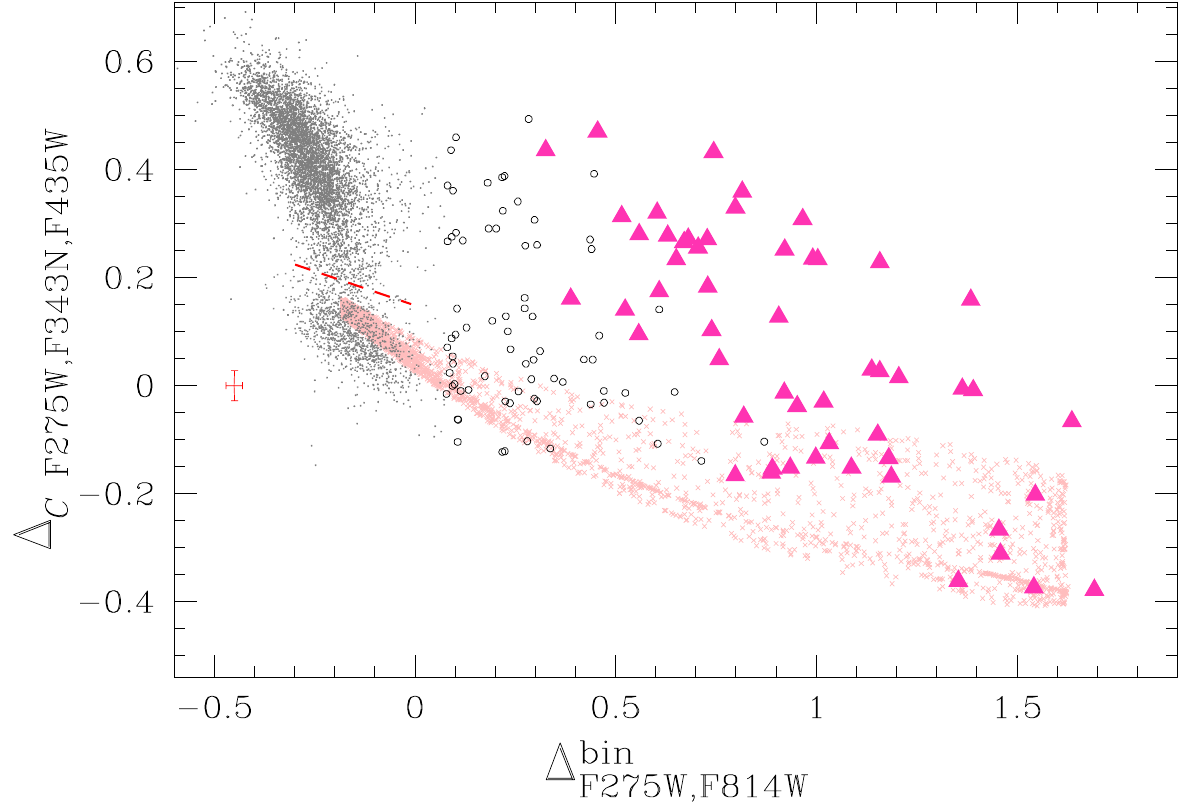}
    \includegraphics[height=.325\textwidth,trim={0.5cm 5.0cm 0.5cm 11cm},clip]{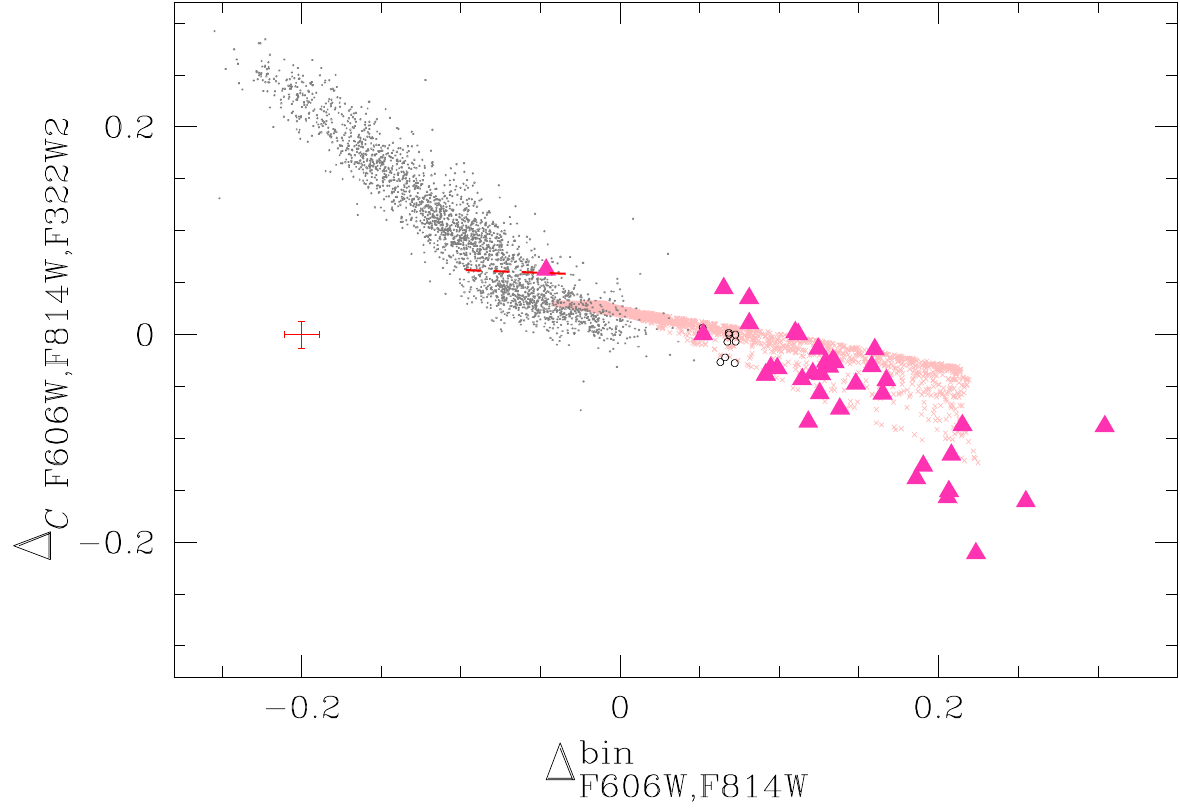}
    \includegraphics[height=.325\textwidth,trim={0.5cm 5.0cm 0.5cm 11cm},clip]{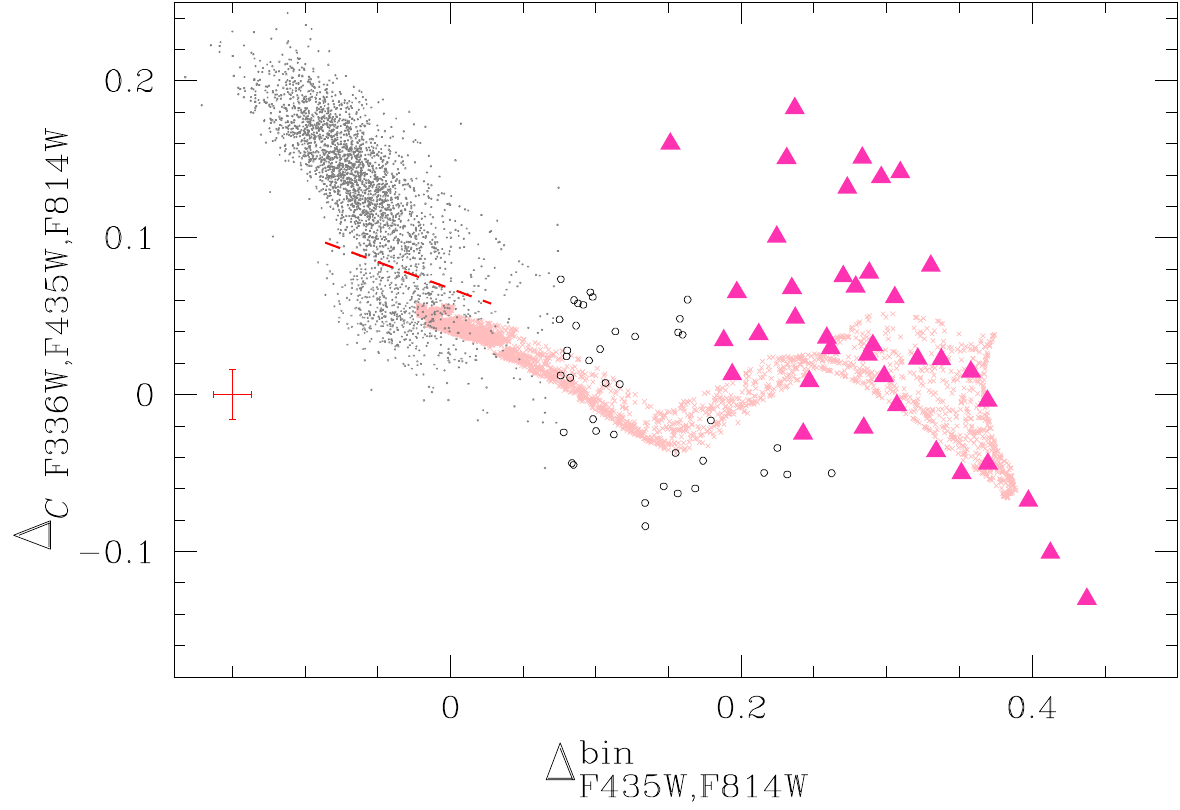}
    \includegraphics[height=.325\textwidth,trim={0.5cm 5.0cm 0.5cm 11cm},clip]{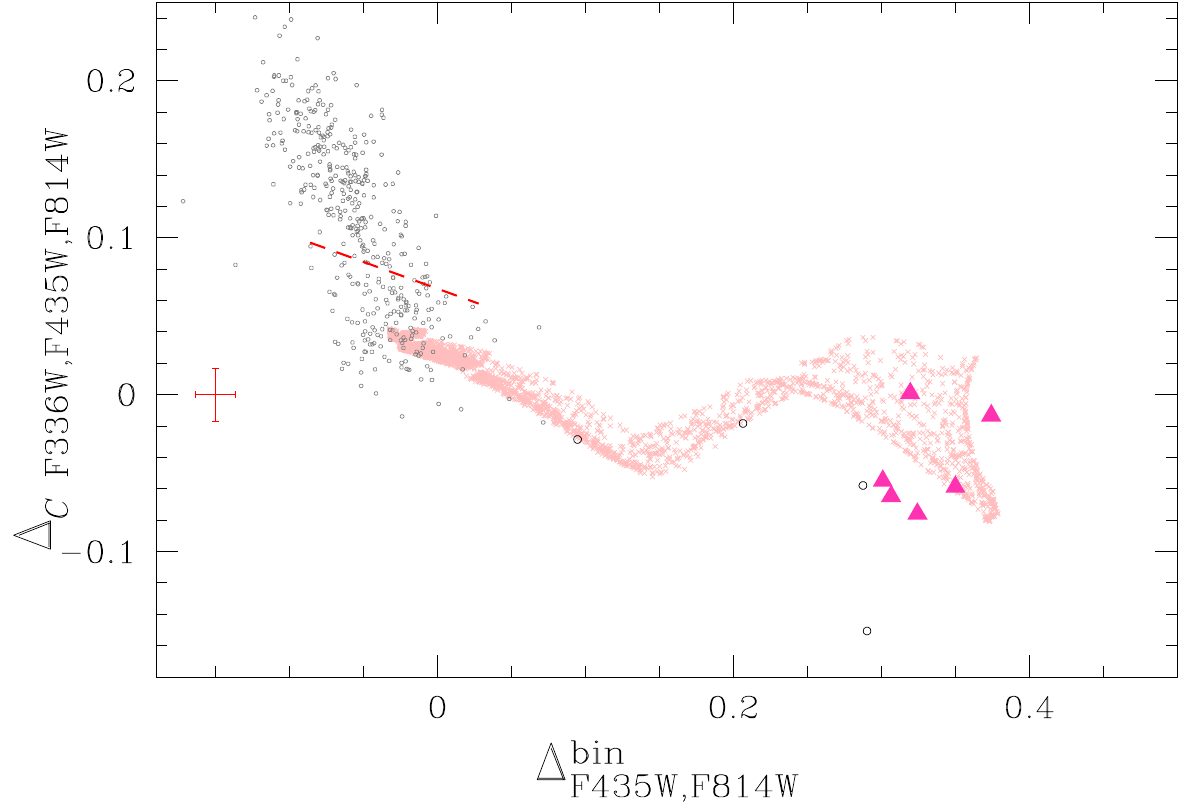}
    \caption{ChMs for stars in the central field (left panels) and in the outer field (right panels). The binaries selected in Figures\,\ref{fig:selbinarie}a, \ref{fig:selbinarieHST}a, \ref{fig:selbinarieCubi1}a, and \ref{fig:selbinarieCubi2}a are represented with large magenta triangles. The remaining candidate binaries are plotted with black circles. The red-dashed lines separate the bulk of single 1P and 2P stars, which are located below and above the lines, respectively. The pink points show simulated 1P binaries.}
    \label{fig:chms}
\end{figure*}

The ChMs of the observed 47\,Tucanae stars are shown in Figure\,\ref{fig:chms}, where we also plot with pink colors the simulated 1P binaries. 
The binary systems studied in Section\,\ref{sec:binaries} are marked with magenta triangles, whereas the remaining probable binaries are represented with black circles. 
The simulated 1P binaries overlap with a small portion of the observed binaries in the central-field ChMs, whereas the majority of both observed and simulated binaries are located in the same region of the outer-field ChM. These observations qualitatively confirm the conclusion of Section\,\ref{sec:binaries} that the central field is populated by both 1P and 2P binaries, whereas 1P binaries are more prevalent in the external regions of 47\,Tucanae.

\subsection{Binaries and multiple populations along the RGB}
\begin{figure*}[htp!]
    \centering
    \includegraphics[height=.6\textwidth,trim={0.5cm 5.0cm 0.5cm 4cm},clip]{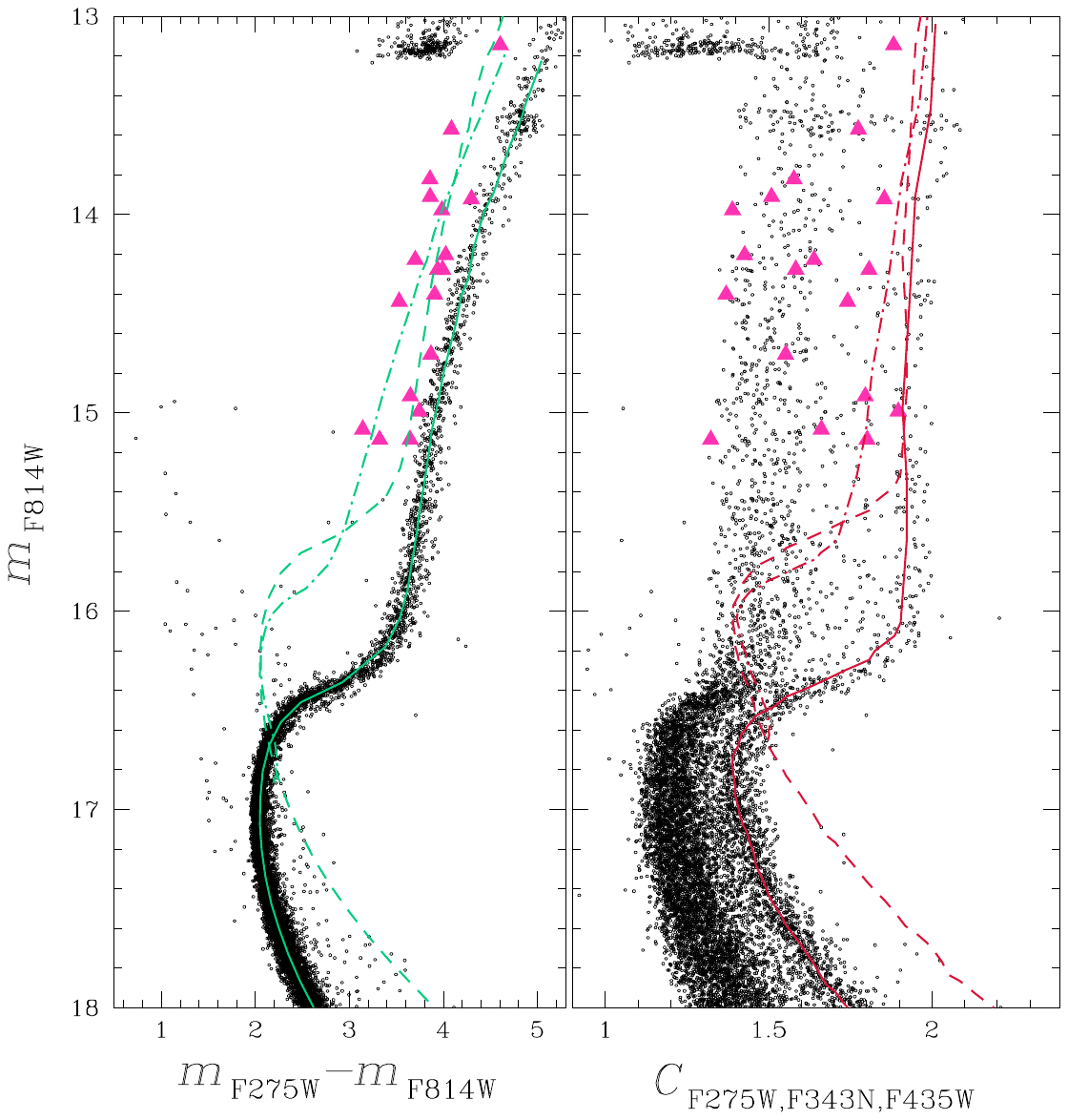}
    \includegraphics[height=.6\textwidth,trim={8.7cm 5.0cm 0.5cm 4cm},clip]{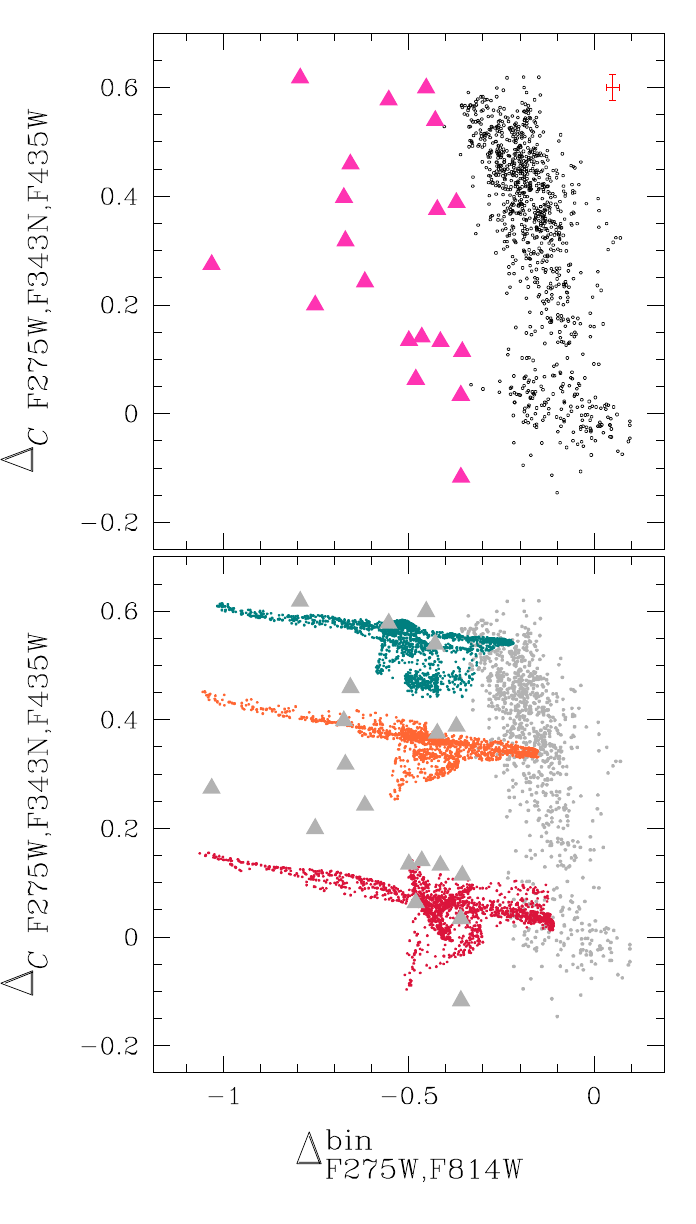}
    \caption{$m_{\rm F814W}$ vs.\,$m_{\rm F275W}-m_{\rm F814W}$ CMD (left) and $m_{\rm F814W}$ vs.\,$C_{\rm F275W,F343N,F435W}$ pseudo-CMD (middle) of stars in the central field zoomed in around the RGB. The solid aqua line plotted in the left-panel CMD is the fiducial line of the MS, SGB, and RGB. The dashed line represents the fiducial of equal-luminosity stars, wheres the fiducial line of binary systems composed of one MS turn off star is plotted with the dot-dashed line. 
     The probable RGB binaries, selected from the $\Delta_{\rm F275W,F343N,F435W}$ vs.\,$\Delta_{\rm F275W,F814W}^{\rm bin}$ ChM shown in the top-right panel, are represented with magenta gray triangles. The bottom-right panel illustrates the simulated ChM for RGB binaries composed of 1P stars (crimson) and 2P stars with intermediate (orange) and extreme (teal) chemical compositions. The gray points reproduce the observed ChM plotted in the top-right panel. }
    \label{fig:RGBs}
\end{figure*}

For completeness, in this subsection we investigate the frequency of 1P and 2P binaries along the RGB. 
To investigate the distribution of binary systems composed of at least one RGB star in the $m_{\rm F814W}$ vs.\,$m_{\rm F275W}-m_{\rm F814W}$ CMD, we show the fiducial line of MS, SGB, and RGB stars (solid aqua line in the left panel of Figure\,\ref{fig:RGBs}) and the fiducial of binary systems composed of pairs of stars with the same luminosity and evolutionary stage (aqua dashed line). We also plot with a dot-dashed line the fiducial line of binaries where one of the two components is located on the MS turn off. The binary systems fainter than $m_{\rm F814W} \sim 13.8$ mag and composed of RGB and MS stars, RGB and SGB stars, or by pairs of RGB stars populated the CMD region limited by the continuous line and the dot-dashed line. In the case of binaries with $m_{\rm F814W} \lesssim 13.8$ mag the blue boundary is the dashed line\footnote{ The fiducial lines shown in the left panel of Figure\,\ref{fig:RGBs} are used to illustrate the CMD region populated by RGB binaries. For simplicity, we derived these fiducials  by considering both 1P and 2P stars. The fiducial of single RGB stars composed by 1P stars only and the fiducials of binary systems that host one 1P RGB star would exhibit slightly redder  $m_{\rm F275W}-m_{\rm F814W}$ colors. Similarly, the corresponding fiducials derived from 2P RGB stars  would be bluer than those plotted in Figure\,\ref{fig:RGBs}. }. This region of the CMD is also populated by evolved blue strugglers.

The middle panel of Figure\,\ref{fig:RGBs} shows the $m_{\rm F814W}$ vs.\,$C_{\rm F275W,F343N,F435W}$ pseudo-CMD of the stars in the central field. The crimson fiducials refer to single 1P stars (solid line), to binaries composed of 1P stars with the same luminosity (dashed line) and binary systems composed of a 1P star located on the MS turn off with a 1P companion (dot-dashed lines). All 1P RGB binaries are located in the area encompassed by these lines.

The $\Delta_{\rm F275W,F343N,F435W}$ vs.\,$\Delta_{\rm F275W,F814W}^{\rm bin}$ ChM is plotted in the top-right panel for RGB stars with $13.0<m_{\rm F814W}<16.0$ mag. This ChM is derived by following the recipe used for the MS but in this case we account for the fact that the fiducial of equal-luminosity RGB binaries is bluer than the RGB boundaries, by using the relation:
 \begin{align}  
 \small
    \Delta_{\rm F275W,F814W}^{\rm bin}=
     \begin{dcases*} \label{eq:1}
             \Delta_{\rm F275W,F814W}, & \text{for $\Delta_{\rm F275W,F814W}\ge -$W },\\ 
        W_{\rm bin} \Bigg{(}\frac{X-X_{\rm fiducial\,B}}{X_{\rm fiducial\,B}-X_{\rm fiducial\,bin}} \Bigg{)-W}, & \text{ for $\Delta_{\rm F275W,F814W}<-$W}
        \end{dcases*}
  \end{align}
\noindent
where the F275W$-$F814W RGB width, $W$, and the quantity $\Delta_{\rm F275W,F814W}$ is defined by \citet[][see their sections 3.1 and 3.2]{milone2017a}, and $W_{\rm bin}$ is the $m_{\rm F275W}-m_{\rm F814W}$ absolute value of the color separation at $m_{\rm F814W}=$14.8 mag between the fiducial line that marks the blue boundary of the MS (fiducial\,B) and the corresponding fiducial line shifted by 0.752 mag to the bright side (fiducial\,bin).  

 As demonstrated by \citet{marino2019a}, RGB binaries exhibit lower values of $\Delta_{\rm F275W,F814W}$ than the bulk of single stars, thus populating the left region of the ChM. We have used magenta triangles to select a sample of 19 probable binaries in the ChM and used the same symbols to represent them in the other panels of Figure\,\ref{fig:RGBs}. As expected, the selected stars populate the region of the $m_{\rm F814W}$ vs.\,$m_{\rm F275W}-m_{\rm F814W}$ CMD occupied by binaries. 
 
 To qualitatively associate binary stars with 1P and 2P stars, we simulated RGB binaries composed of 1P RGB stars, and 2P RGB stars with intermediate and extreme chemical compositions. The simulated binaries are represented with crimson, orange, and teal colors, respectively, in the ChM plotted in the bottom-right panel of Figure\,\ref{fig:RGBs} and are superimposed on the observed RGB stars (gray points). We notice that half a dozen binaries only are consistent with 1P binaries, whereas the remaining selected binaries are probably composed of 2P stars or mixed.   A similar conclusion is provided by the visual inspection of the $m_{\rm F814W}$ vs.\,$C_{\rm F275W,F343N,F435W}$ pseudo-CMD plotted in the middle panel of Figure\,\ref{fig:RGBs}. 
 A quantitative determination of the frequency of binaries among 1P and 2P RGB stars is challenged by the fact that the selected candidate binaries could comprise evolved blue stragglers, for which accurate models are not available to us. Nevertheless, the comparison between the observed ChM of RGB stars in 47\,Tucanae and the simulated ChMs of 1P and 2P binaries corroborates the conclusion derived from MS stars that both 1P and 2P binaries populate the central region of 47\,Tucanae. 
 
\begin{figure}[htp!]
    \centering
    \includegraphics[height=.3\textwidth,trim={0.5cm 5.0cm 0.5cm 12.2cm},clip]{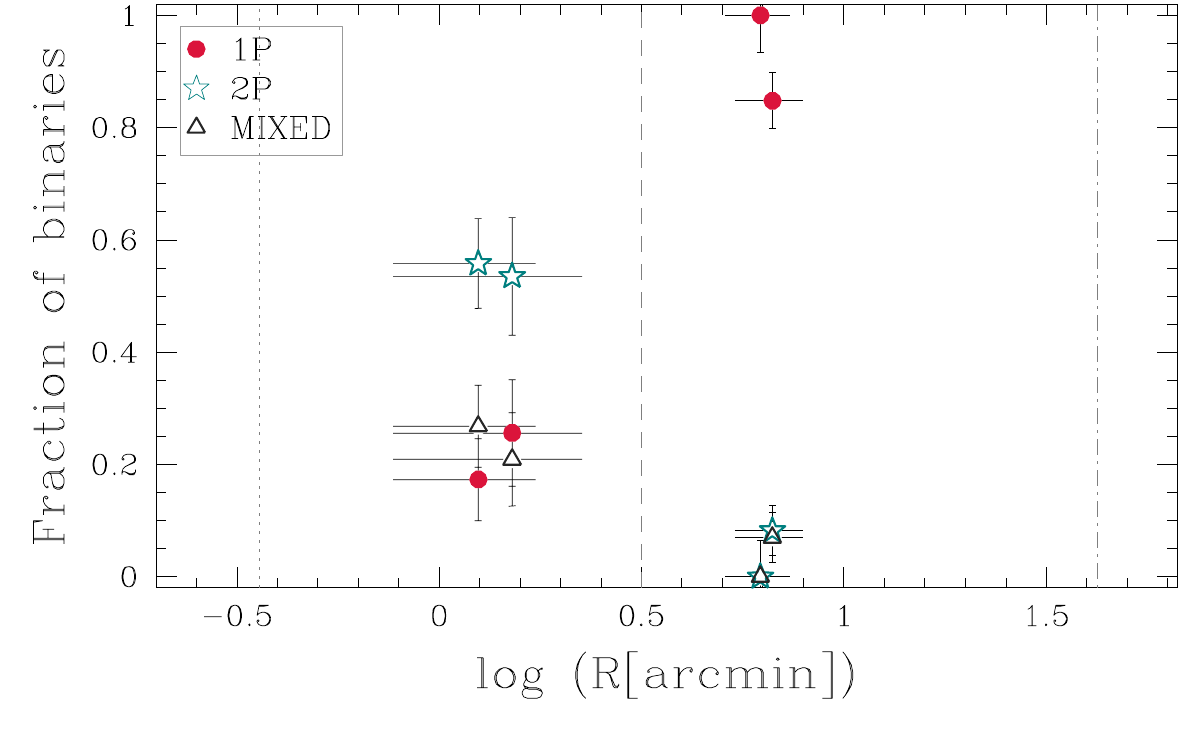}
    \includegraphics[height=.3\textwidth,trim={0.5cm 5.0cm 0.5cm 12.2cm},clip]{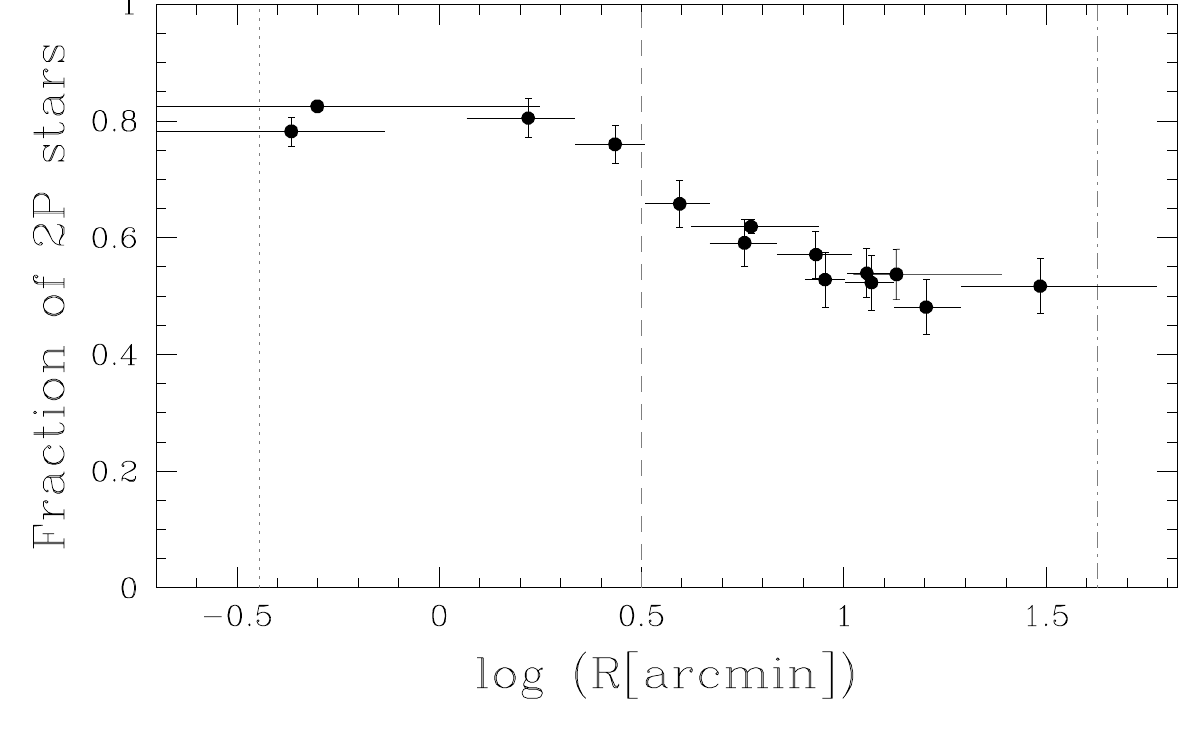}
     \caption{  {\it Top.} Fractions of 1P binaries (dots), 2P binaries (starred symbols), and mixed binaries (triangles) derived in this paper against the logarithm of the radial distance from the center of 47\,Tucanae.  {\it Bottom.} Fraction of 2P stars as a function of radial distance, in logarithm scale, from literature works \citep{milone2017a, dondoglio2021a, marino2024a, mehta2025a}.     
     The vertical dotted, dashed, and dashed-dotted lines mark the core, half-light, and tidal radius, respectively, from the 2010 edition of the \cite{harris1996a} catalog.}
    \label{fig:RDs}
\end{figure}

\section{Summary and discussion}\label{sec:summary}
In this paper, we used multi-band {\it HST} and {\it JWST} photometry of 47\,Tucanae to estimate the fraction of binaries among 1P and 2P stars. 
   In contrast with the previous works on other GCs, which are limited either to the cluster center or the external regions \citep{dorazi2010a, lucatello2015a, dalessandro2018a, marino2019a, kamann2020a, milone2020a}, we homogeneously analysed two distinct fields with different radial distances from the center of 47\,Tucanae. 

   We used the method by \cite{milone2020a}, which consists of three main steps, including i) the identification of 1P and 2P stars, ii) the selection of binary systems composed of stars with similar luminosities, and iii) the analysis of pseudo-colors of the selected binaries, which are indicative of their chemical composition. 
 
        In the central field, we took advantage of the $\Delta_{C {\rm F275W,F343N,F435W}}$ vs.\,$\Delta_{\rm F275W,F814W}$ ChM to identify the bulk of 1P and 2P stars along the upper MS, together with five sub-groups of 2P$_{\alpha - \epsilon}$ stars. 
        We identified a sample of MS-MS binaries by using the pseudo-CMD $m_{\rm F814W}$ vs.\,($m_{\rm F336W}-m_{\rm F814W}$)$-$0.40($m_{\rm F606W}-m_{\rm F814W}$)), where the fiducial lines of the multiple stellar populations are nearly coincident. 
     Finally, we analyzed the $C_{\rm F275W,F343N,F435W}$ pseudo-color distribution of the observed binary stars and compared it to the distributions from a large set of simulated stellar populations. These simulations encompassed various combinations of 1P, 2P, and mixed stars.
     We find that about 18\% and 56\% of the studied binaries are composed of 1P and 2P stars, respectively. The remaining $\sim 26$\% of binaries are probable mixed stellar systems.

        In the external field, we analyzed both M-dwarf stars. The 1P and 2P stars are identified from the $\Delta_{C {\rm F606W,F814W,F322W2}}$ vs.\,$\Delta_{\rm F606W,F814W}$ ChM, whereas the binaries are selected from the $m_{\rm F814W}$ vs.\,($m_{\rm F606W}-m_{\rm F160W}$)$-$0.55($m_{\rm F606W}-m_{\rm F814W}$) pseudo CMD. In this case, the fraction of binaries among the multiple populations is inferred by comparing the $\Delta_{C {\rm F606W,F814W,F322W2}}$ pseudo-color of the selected binaries with that of a large sample of simulated binaries. Most binaries, $\sim 85$\%, are composed of 1P stars. 
        These results are illustrated in the top panel of Figure\,\ref{fig:RDs} and are confirmed by the analysis of bright MS stars in the $m_{\rm F814W}$ vs.\,$C_{\rm F336W,F435W,F814W}$ pseudo-CMD in both the inner and outer fields. 
        
        As illustrated in the bottom panel of Figure\,\ref{fig:RDs}, the multiple stellar populations of 47\,Tucanae exhibit different radial distributions, with 2P stars being significantly more centrally concentrated than the 1P \citep[e.g.][]{milone2012a, cordero2014a, dondoglio2021a, lee2022a, mehta2025a}. In particular, the fraction of 1P stars in the cluster center and in the external field studied in this paper is $\sim$20\% and 40\%, respectively \citep{milone2017a, marino2024a}.

        The incidence of 1P binaries can be estimated as the ratio between the fractions of 1P binaries and the fractions of 1P stars \citep[from][]{milone2012a, marino2024a} and the analogous relation can be adopted for the incidence of 2P binaries. 
        Hence, in the cluster center, the incidence of 1P binaries is slightly ($\sim$1.3 times) higher than that of 2P binaries. In contrast, the outer regions predominantly feature 1P binaries.

        The present-day fractions of binaries among 1P and 2P stars provide information on the origin and the dynamic evolution of multiple stellar populations in GCs. 
        Binary systems have been widely investigated in the context of GC formation scenarios where the 2P stars formed in a high-density environment in the innermost region of a more-extended 1P. Such configuration is adopted in various scenarios for the formation of 2P stars \citep[e.g.][]{dercole2008a, dercole2012a, bekki2010a, lacchin2022a} and consistent with the expected initial configuration of various scenarios \citep[e.g.\,][]{bastian2013a, gieles2018a}.

        Studies based on analytic calculations and N-body simulations suggest that the exact values of the current fractions of 1P and 2P binaries are influenced by several factors. These include the initial numbers and configurations of 1P and 2P stars, the evolution and disruption of binary systems, and the changes in the spatial distributions of 1P and 2P single and binary stars over time.
        Nevertheless, these simulations show that, due to the dense environment, 2P binaries evolve and are disrupted at a much higher rate than 1P binaries. 
        Consequently, the present-day 2P population is expected to have a lower global binary incidence compared to the 1P population \citep{vesperini2011a, hong2015a, hong2016a, hong2019a}. These studies, which assume different initial radial distributions of 1P and 2P stars with the 2P initially more centrally concentrated, predict radial variations in the incidence of 1P and 2P binaries, with the most pronounced differences observed in the outermost regions of the cluster \citep[e.g.][]{hong2016a, milone2020a}.
    
        The evidence of a radial gradient in the incidence of 1P and 2P binaries in 47\,Tucanae, and the fact that the external regions are mostly populated by 1P binaries are in general agreement with the predictions of those theoretical studies, and corroborate the possibility that 2P stars formed in a dense subsystem in the innermost cluster regions. Similarly, the finding of mixed binary systems composed of one 1P and one 2P star is qualitatively consistent with the findings of the simulations by \citet{hong2015a, hong2016a} who predicted that, due to stellar encounters, one of the binary-system components can be replaced by one of the interacting stars of a different population, thus producing a mixed binary system.

\begin{acknowledgements}
We thank the anonymous referee for their valuable suggestions, which have improved the manuscript.
This work has been funded by the European Union – NextGenerationEU RRF M4C2 1.1 (PRIN 2022 2022MMEB9W: “Understanding the formation of
globular clusters with their multiple stellar generations”, CUP C53D23001200006). (PI Anna F.\,Marino), 
from INAF Research GTO-Grant Normal RSN2-1.05.12.05.10 -  (ref. Anna F. Marino) of the "Bando INAF per il Finanziamento della Ricerca Fondamentale 2022", and from the European Union’s Horizon 2020 research and innovation programme under the Marie Skłodowska-Curie Grant Agreement No. 101034319 and from the European Union – NextGenerationEU (beneficiary: T. Ziliotto). SJ acknowledges support from the National Research Foundation of Korea (2022R1A2C3002992, 2022R1A6A1A03053472). J.-W. Lee acknowledges financial support from the Basic Science Research Program (grant No. 2019R1A2C2086290) through the National Research Foundation of Korea.
\end{acknowledgements}

\bibliography{aanda}
\end{document}